\documentclass[a4paper,10pt]{article}
\usepackage[sectionbib]{natbib}
\pdfoutput=1
\usepackage[utf8]{inputenc}
\usepackage{graphicx}		
\usepackage{amsmath}		
\usepackage{setspace}		
\usepackage[a4paper,tmargin=2.5cm,bmargin=3cm,rmargin=2cm,lmargin=2cm]{geometry} 
\usepackage[left]{lineno}	
\usepackage{color}		
\usepackage{float}
\usepackage{multirow}
\usepackage{textcomp}
\usepackage{tabulary}
\usepackage{bm}
\usepackage{xr}
\usepackage{url}

\providecommand{\fst}[1]{\ensuremath{F_{\text{ST}}}}
\title{A new $F_{\text{ST}}$-based method to uncover local adaptation using environmental variables.}
\author{Pierre de Villemereuil\footnotemark[1], \& Oscar E. Gaggiotti\footnotemark[1]~\footnotemark[2]\\
\small{\footnotemark[1]~: Universit\'{e} Jospeh Fourier, Centre National de la Recherche Scientifique,}\\
\small{LECA, UMR 5553, 2233 rue de la piscine, 38400 Saint Martin d'H\`{e}res, France}\\
\small{\footnotemark[2]~Scottish Oceans Institute, University of St Andrews,}\\
\small{Fife, KY16 8LB, United Kingdom}
}
\date{}

\begin{document}

\maketitle


\textbf{Corresponding author:} Pierre de Villemereuil, E-mail: bonamy@horus.ens.fr

 \begin{abstract}
 \textbullet~Genome-scan methods are used for screening genome-wide patterns of 
 DNA polymorphism to detect signatures of positive selection. There are two 
 main types of methods: \textit{(i)} ``outlier'' detection methods based on 
 \fst{} that detect loci with high differentiation compared to the rest of the 
 genomes, and \textit{(ii)} environmental association methods that test the 
 association between allele frequencies and environmental variables.\\
 \textbullet~We present a new \fst{}-based genome-scan method, BayeScEnv, which 
 incorporates environmental information in the form of ``environmental 
 differentiation''. It is based on the \textit{F} model, but, as opposed to 
 existing approaches, it considers two locus-specific effects; one due to 
 divergent selection, and another one due to various other processes different from 
 local adaptation (e.g. range expansions, differences in mutation rates across 
 loci or background selection). The method was developped in C++ and is 
 avaible at \url{http://github.com/devillemereuil/bayescenv}.\\
 \textbullet~Simulation studies shows that our method has a much lower false 
 positive rate than an existing \fst{}-based method, BayeScan, under a wide 
 range of demographic scenarios. Although it has lower power, it leads to a 
 better compromise between power and false positive rate.\\
 \textbullet~We apply our method  to human and salmon datasets and show that it can be used 
 successfully to study local adaptation. We discuss its scope and compare its mechanics to other existing methods.
 \end{abstract}

\section*{Introduction}

One of the most important aims of population genomics \citep{luikart_power_2003} is to 
uncover signatures of selection in genomes of non model species. Of special interest is the process
of local adaptation, whereby populations experiencing different environmental conditions undergo 
adaptive, selective pressures specific to their local habitat. As a result, populations evolve traits that provide an advantage 
in their local environment. Many experimental approaches focused on potentially adaptive traits have been developed to test for local adaptation 
\citep[reviewed in][]{blanquart_practical_2013}, but only recently it has become possible to make inferences about 
the genomic regions involved in local adaptation processes. Indeed, the advent of next generation sequencing \citep[NGS,][]{shendure_next-generation_2008} has fostered 
the development of so-called genome-scan methods aimed at identifying regions of the genome subject to selection. These methods
are now widely used in studies of local adaptation \citep{faria_advances_2014}.

There are two main types of genome-scan methods. The first type detects `outlier' loci using locus-specific \fst{} estimates, 
which are compared to either an empirical distribution \citep{akey_interrogating_2002}, or to a distribution expected under a 
neutral model of evolution \citep{beaumont_identifying_2004,foll_genome-scan_2008}. The rationale behind these methods is that 
local adaptation leads to strong genetic differentiation between populations, but only at the selected loci (or marker loci 
linked to them). Thus, loci with very high \fst{} compared to the rest of the genome are suspected to be under strong local 
adaptation and are referred to as outliers. The outlier approach was further extended to statistics akin to \fst{}
\citep{bonhomme_detecting_2010,gunther_robust_2013},
and also to other unrelated statistics \citep{duforet-frebourg_genome_2014}. One limitation of these methods is that they are not 
designed to test hypotheses about the environmental factors underlying the selective pressure.


A second type of methods focuses on environmental variables and aims at associating patterns of allele frequency to 
environmental gradients. The rationale is that selective pressures should create associations between allele frequencies at the 
selected loci and the causal environmental variables \citep{coop_using_2010}. In the presence of 
population structure, performing a simple linear regression would be an error-prone approach 
\citep{de_mita_detecting_2013,de_villemereuil_genome_2014}. Instead, existing methods account for population
structure by modelling the allele frequency covariation across populations \citep{coop_using_2010,frichot_testing_2013,guillot_detecting_2014}. 
One disadvantage of these approaches is that the parameters that capture the 
effect of demographic history on genetic differentiation do not have a clear biological interpretation, which in turn makes 
the rejection of the null model hard to interpret in terms of detection of local adaptation. We note that although the elements
of the covariance matrix estimated by \citet{coop_using_2010} could in principle be interpreted as parametric estimates of the 
pairwise and population-specific \fst{}, this is only true when levels of genetic drift are low 
\citep{nicholson_assessing_2002}. 

It is important to note that, regardless of the type of genome-scan method under consideration, processes other than local 
adaptation might be responsible for the observed spatial patterns in allele frequency or \fst{}. These include demographic processes \citep[e.g. allele surfing;][]{edmonds_mutations_2004}, large differences
in mutation rate across loci \citep{edelaar_comparisons_2011}, hybrid incompatibility following secondary contact 
\citep{kruuk_comparison_1999} and background selection \citep{charlesworth_measures_1998}.
It is therefore possible that some of the loci identified as outliers are in fact false positives. Accounting for processes 
other than selection would require introducing parameters that could appropriately capture the effect of these other processes. 

Here, we present a method that incorporates features of the two types of genome-scans described above. The objective is to better 
discriminate between true and false genetic signatures of local adaptation, and simultaneously allow inferences about 
the environmental factors underlying selective pressures. More precisely, our method is based on the Bayesian approach first 
proposed by \citet{beaumont_identifying_2004} and later extended by \citet{foll_genome-scan_2008}. The original formulation 
considers population- and locus-specific \fst{}'s, which are described by a logistic regression model with three parameters: a 
locus-specific term, $\alpha_i$, that captures the effect of mutation and some forms of selection, a population-specific term, 
$\beta_j$, that captures demographic effects (e.g. $N_e$ and migration) and a locus-by-population interaction term, 
$\gamma_{ij}$, that reflects the effect of local adaptation. The estimation of the first two terms benefits from sharing 
information across loci or populations, but this is not the case for the interaction term, which is therefore poorly estimated 
(\citealp{beaumont_identifying_2004}, but see \citealp{riebler_bayesian_2008}). In practice signatures of local adaptation are therefore inferred from the locus-specific 
effects ($\alpha_i$) under the assumption that large positive values reflect adaptive selection. The implicit assumption is that 
background selection and mutation should not have much of an effect on this regression term. In order to relax this assumption 
and to better estimate the interaction term we introduce environmental data so that $\gamma_{ij}=g_{i}E_{j}$, where $E_j$ is 
the ``environmental differentiation'' observed in population $j$ and $g_{i}$ is a locus-specific regression coefficient. 
In what follows, we first describe in detail the probabilistic model underlying our Bayesian approach. We then evaluate its performance using simulated data and present an application using 
human and salmon datasets. Finally, we discuss the scope of our method and compare it with other existing genome-scan approaches.

\section*{Statistical model}

\subsection*{Modelling allele frequencies using the $F$ model}
Our new genome-scan approach is based on the $F$ model 
\citep{beaumont_identifying_2004,foll_genome-scan_2008} and extends the software BayeScan 
\citep{foll_genome-scan_2008} by incorporating environmental data so as to explicitly consider local adaptation 
scenarios. Full details of the $F$ model are given by \citet{gaggiotti_quantifying_2010}, so here we only provide a 
brief description. The core assumptions of the $F$ model is that all populations share a common pool of migrants, but 
that their effective sizes and immigration rates are population-specific. Thus, population structure at each locus 
is described by local \fst{}'s that measure genetic differentiation between each local population and the migrant 
pool.

The $F$ model uses the multinomial-Dirichlet likelihood for the allele counts 
$\bm{a_{ij}}=(a_{i,j,1},\dots,a_{i,j,K_{i}})$ at locus $i$ within population $j$ (where $K_{i}$ is the number of 
distinct alleles at locus $i$) with parameters given by the migrant pool allele frequencies, 
$\bm{f_i}=(f_{i,1},\dots,f_{i,K_{i}})$, and a population- and locus-specific parameter of similarity, 
$\theta_{ij}=\dfrac{1-F_{\text{ST}}^{ij}}{F_{\text{ST}}^{ij}}$:
\begin{equation}\label{multD}
 \bm{a_{ij}} \sim \text{\it multDir}(\theta_{ij}f_{i,1},\dots,\theta_{ij}f_{i,K_{i}}),
\end{equation}
where $\text{\it multDir}$ stands for the multinomial-Dirichlet distribution.\\
Although, for the sake of simplicity, we only present here the formulation for co-dominant data, the software 
implementing our approach also allows for dominant data (e.g. AFLP markers) using the same probabilistic 
model as \citet{foll_genome-scan_2008}. Note finally that, for bi-allelic co-dominant markers (e.g. SNP markers), 
the likelihood reduces to a beta-binomial model. 

\subsection*{Alternative models to explain population structure}
Our purpose is to better discriminate between true signals of local adaptation and spurious signals left by other processes.
Therefore, we assume that genetic differentiation at individual loci is influenced by three type of effects: \textit{(i)} genome-wide effects due
to demography, \textit{(ii)} a locus-specific effect due to local adaptation caused by the focal environmental variable, and \textit{(iii)} locus-specific 
effects unrelated to the focal environmental variable. 
Although in principle one could consider all seven alternative model that can be constructed with different combinations of these 
three effects, most of them would not have any biological meaning. For example, all models should include genome-wide effects 
associated with genetic drift. Additionally, we do not consider the two types of locus-specific effects simultaneously in a full model.
The reason is that the statistical (and hence biological) interpretation of  $\alpha_i$ will depend on whether or not the parameter $g_i$ is included in the model.


This can render the algorithms overly complicated, especially during the pilot runs (see below).

Thus, we focus on three alternative models to explain the genetic structuring at individual loci.

\paragraph{Null model of population structure}
Under the null hypothesis that all loci are neutral, the local differentiation parameter $F_{\text{ST}}^{ij}$ will 
be driven only by local population demography and, hence, should be common to all loci:
\begin{equation}\label{eq_beta}
 \log\left(\frac{F_{\text{ST}}^{ij}}{1-F_{\text{ST}}^{ij}}\right) = \log\left(\frac{1}{\theta_{ij}}\right) = \beta_{j}.
\end{equation}


A high $\beta_{j}$ value means that the population $j$ is strongly differentiated from the pool of migrants. 
This could be due to a lack of immigration from the other populations, 
a reduced effective size, or a particular spatial structure.

\paragraph{Alternative model of local adaptation}
In this model, we focus on a particular signature left by a process of local adaptation. If selection is driven by a 
putative environmental factor, we expect that genetic differentiation for the locus or loci under selection will be 
stronger than expected under neutrality for populations with strong environmental differentiation. Any measure of 
distance between the environmental value of population $j$ and the average environment could serve as a measure of 
 differentiation. For the sake of simplicity, we here only consider the absolute value (i.e. Manhattan distance). 
Its advantage is that it does not over-state the importance of outlier environmental values. Furthermore, 
in order to facilitate the calibration of prior distributions, we only consider standardised environmental values
(i.e. with zero mean and unit variance).\\
To model the effect of local adaptation on locus $i$, we consider the impact of environmental differentiation 
$E_{j}$ of population $j$ on the locus, we thus modify Eq. \ref{eq_beta} as follows:
\begin{equation}\label{eq_g}
 \log\left(\frac{F_{\text{ST}}^{ij}}{1-F_{\text{ST}}^{ij}}\right) = \beta_{j} + g_{i}E_{j},
\end{equation}
where $g_{i}$ quantifies the sensitivity of locus $i$ to the environmental differentiation.

\paragraph{Alternative model of locus-specific effect}
Local adaptation with respect to the focal environmental variable is not the only evolutionary phenomenon that could lead to 
departures from the neutral model. Other phenomena that could produce such locus-specific effects
include local adaptation due to other unknown factors, large differences in mutation rate across loci, the so-called allele surfing
phenomenon \citep{edmonds_mutations_2004} and background 
selection \citep{charlesworth_background_2013}. 

This is accounted for by using the following parametrisation for local differentiation:
\begin{equation}\label{eq_alpha}
 \log\left(\frac{F_{\text{ST}}^{ij}}{1-F_{\text{ST}}^{ij}}\right) = \alpha_{i}+\beta_{j}.
\end{equation}
The main advantage of implementing both of the above alternative models is that we can distinguish between departures 
from the neutral model of unknown origin (using Eq. \ref{eq_alpha}) and departures due to local adaptation
caused by a particular environmental factor (using Eq. \ref{eq_g}).

\begin{figure}
 \centering
 \includegraphics[width=0.5\textwidth]{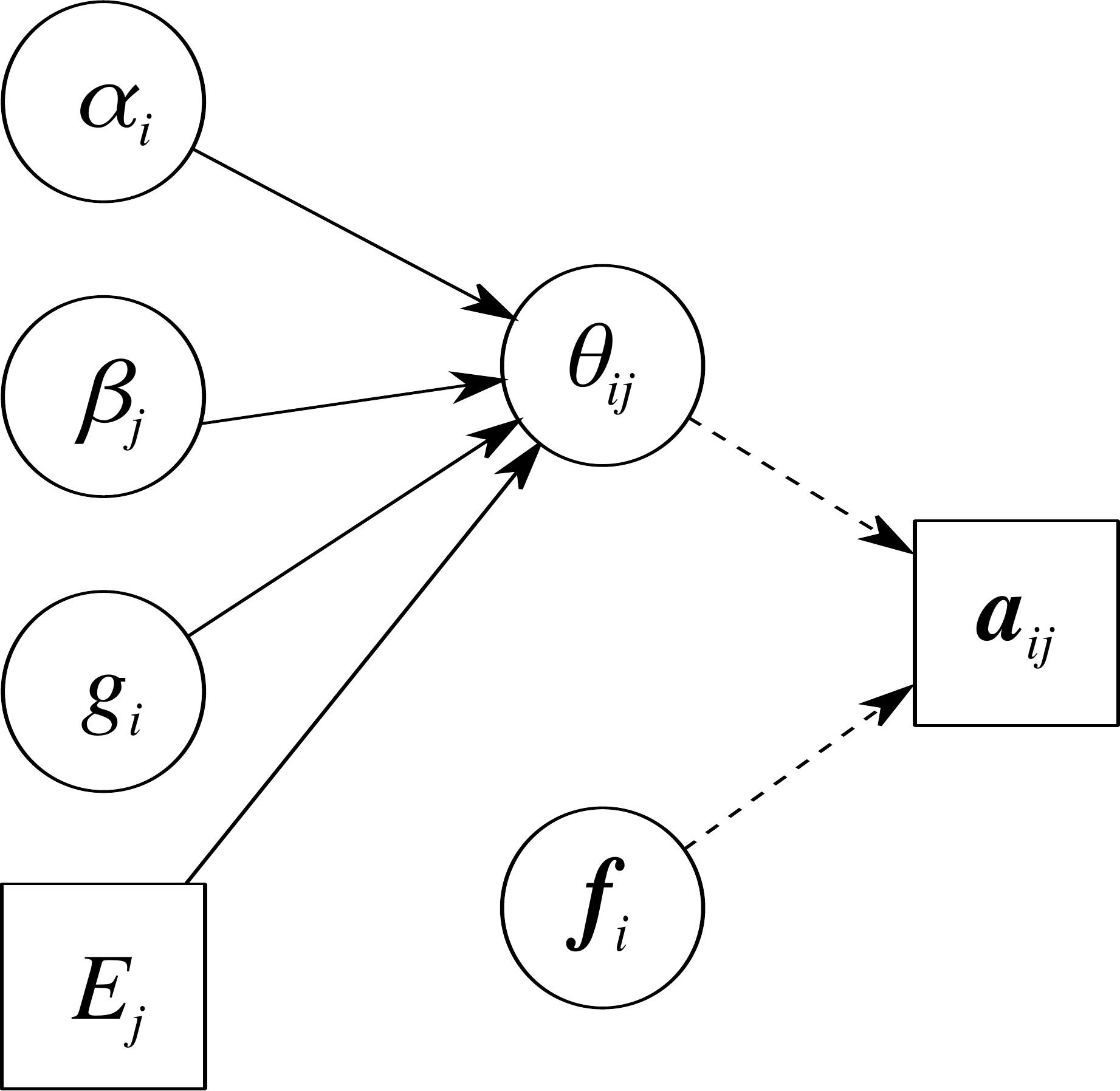}
 \caption{Directed Acyclic Graph (DAG) of the model. Squared nodes denote know quantities ($E$ for 
 environmental data, and $A$ for genetic marker data). Circled nodes denote unknown parameters. Plain arrows 
 stand for deterministic relationships, and dashed arrows stand for stochastic relationships.}
 \label{fig_dag} 
\end{figure}

\section*{Material and Methods}

\subsection*{Implementation of the statistical model}

Our method uses two types of data: \textit{(i)} the allele counts $\bm{a}$ for each locus in each 
population sample, and \textit{(ii)} observed values $\bm{E}$ of an environmental variable (one value per  
population), which are transformed into environmental differentiation using an appropriate function. We chose the 
absolute-value distance, because it allows to weigh down the effect of outlier (i.e. strongly 
differentiated) environmental values and, therefore, makes the method more conservative. 
Note that measuring an environmental distance requires to define a reference. The most
natural reference would be the average of the environmental values, but this would not be always the case (see the example of  
adaptation to altitude in humans presented below). Also, it is strongly advised to standardise the environmental values by dividing by the
standard deviation, in order to avoid effect size issues regarding the inference of the parameter $g$.\\
As stated in the previous section, there are three alternative models:
\begin{description}
 \item[M1] Neutral model: $\beta_{j}$,
\item[M2] Local adaptation model with environmental differentiation $E_{j}$: $\beta_{j} + g_{i}E_{j}$,
\item[M3] Locus-specific model: $\alpha_{i}+\beta_{j}$.
\end{description}
All three models were implemented using an RJMCMC algorithm \citep{green_reversible_1995}. In order to propose relevant 
values for new parameters during the jumps, the RJMCMC is preceded by pilot runs. These are aimed at both calibrating the 
MCMC proposals to reach efficient acceptance rates, and approximating the posterior distribution of parameters, 
as proposed by \citet{brooks_markov_1998} and already implemented in BayeScan \citep{foll_genome-scan_2008}. 
Our code is based on the source code of BayeScan 2.1 
and is written in C++. The source and binaries are available at \url{https://github.com/devillemereuil/bayescenv}.\\
Our prior belief in the three models is described by two parameters: the probability $\pi$ of moving away from the neutral 
model and the preference $p$ for \textbf{M3} against \textbf{M2} as alternative models. We can calculate the prior 
probability for each model as:
\begin{equation}
 \begin{array}{rcl}
 P(\text{\textbf{M1}}) & = & 1-\pi, \\
 P(\text{\textbf{M2}}) & = & \pi (1-p),\\
 P(\text{\textbf{M3}}) & = & \pi p.\\
 \end{array}
\end{equation} 
The details of the mathematical calculation of transition between models can be found in the Supplementary Material.
Pilot studies showed that using values of $p$ above 0.5 yielded extremely conservative results.
\\
We used a uniform Dirichlet prior for the allele frequencies $\bm{f}_{i} \sim \text{\it Dir}(1,\dots,1)$. The priors 
for the hyperparameters $\alpha$ and $\beta$, were Normal with mean -1 and variance 1.
Since under a local adaptation scenario the parameter $g$ is only
expected to be positive, it was assigned a uniform prior between 0 and 10.

Our method outputs posterior error probabilities and $q$-values, which are test statistics related to the False Discovery Rate (FDR) 
 \citep{storey_direct_2002,kall_posterior_2008}. Contrary to the commonly used 
False Positive Rate (FPR), which is the probability of declaring a locus as positive given that it is actually 
neutral, the FDR is the proportion of the positive results that are in fact false positives, and is more appropriate for
multiple testing \citep{kall_posterior_2008}. See the Supplementary Information (SI) for more details.

\subsection*{Simulation analysis}
We performed a simulation study to evaluate the performance of our method and compare it with that of BayeScan \citep{foll_genome-scan_2008}. 
We modelled 16 populations each with 500 individuals genotyped at 5,000 loci, 
among which one (monogenic scenario) or 50 (polygenic scenario) were under selection. We modelled three kinds of population 
structure: \textit{(i)} a classical island model (IM), \textit{(ii)} a one-dimension stepping-stone (SS) model and 
\textit{(iii)} a hierarchically structured (HS) model.\\
The genome was composed of 5,000 bi-allelic SNPs spread along 10 chromosomes.
The loci under selection, one for the monogenic case and 50 for the polygenic case, were randomly distributed across the genome. 
Since all markers were independently initialised, our simulations yielded negligible linkage disequilibrium. Consequently, we considered as 
true positives only the loci subject to selection. For the IM and SS scenarios,
we directly initialised all 16 populations. For the HS scenario, we initialised the ancestral population, which, following successive and temporally spaced-out 
fission events, gave rise to 2, 4, \dots\ , 16 populations. This hierarchical structure is reinforced by 
preferential migration between related populations. More details regarding migration and population history are available 
in the SI. This model is very close to that used by \citet{de_villemereuil_genome_2014}. 
It should be particularly difficult for our method, because all populations are equally differentiated (i.e. 
the $\beta_{j}$ parameters are expected to be roughly the same across populations), but a phylo-geographic covariance 
exists between related populations, which is not explicitly accounted for by our probabilistic model. Information
regarding the environmental gradient and the fitness function are available in the SI.\\
The simulations were performed using the SimuPOP Python library \citep{peng_simupop:_2005} and the scripts are available online 
in the data section. Our simulated datasets were analysed using our C++ code and version 2.1 of BayeScan 
\citep{foll_genome-scan_2008}.\\
We generated 100  datasets for each
scenario and computed the realised FDR, FPR and power yielded by BayeScan and our new 
environmental method (BayeScEnv). For the latter, we also compared several parametrisations using a prior probability $\pi$ 
of jumping away from the neutral model of $0.1$ (equivalent to the default prior odds used by BayeScan, which is 10) or  $0.5$, 
as well as a preference for the locus-specific model $p$ of 0.5 (environmental and locus-specific models are equiprobable) or 0 
(the locus-specific model is forbidden and only the environmental model is tested against the neutral one).

\subsection*{HGDP SNP data analysis}
In order to test our new method against a real dataset, we focused on 26 Asian populations from the Human 
Genome Diversity Panel (HGDP) SNP Genotyping data (\url{http://www.hagsc.org/hgdp/files.html}). This data set 
consists of 660,918 SNP markers genotyped using Illumina 650Y arrays.
After cleaning the dataset from mitochondrial and sex-linked markers, we removed all markers with minor 
allele frequency below 5\%. This left us with a total of 446,117 SNPs. For all populations, we obtained the following environmental 
variables from the BIOCLIM database (\url{http://worldclim.org/bioclim}): mean annual temperature, precipitation,  
and altitudinal data. We ran separate BayeScEnv analysis for each variable and compared the results with BayeScan 
(which doesn't use environmental variables). After standardisation of the environmental variables, we computed environmental 
differentiation from the mean for temperature and precipitation, and from the sea level for elevation. Gene ontology 
enrichment tests for the detected genes were performed using the ``SNP mode'' of the Gowinda software 
\citep{kofler_gowinda:_2012}.
The prior odds for BayeScan was 10 for this analysis. BayeScEnv prior parameters for this analysis were $\pi=0.1$ and $p=0.5$.

\subsection*{Atlantic salmon data analysis}
We downloaded genetic markers and environmental data for Atlantic salmon (\textit{Salmo salmar}) from the Dryad database 
\citep{bourret_data_2014}. The data included 3118 SNP markers obtained using Expressed Sequence Tags (EST) 
and Genome Complexity Reduction \citep[GCR, see][]{bourret_snp-array_2013}. The dataset consists of 23 populations from North 
American coasts. The dataset was cleaned by removing markers with minor allele frequency below 5\%, leading to a final 
dataset of 2078 markers. Environmental data comprised 53 variables, including information relative to local temperature, precipitation,
river and geological properties. A PCA was performed. For further analyses, we retained the 3 first axes which 
were respectively correlated with temperature, precipitations and river properties, as indicated by 
\citet{bourret_landscape_2013}. Since PCA scores are already standardised values, we used these variables as such.
For the sake of simplicity, we called the first axis ``temperature'', the second ``precipitation'' and the third
``river properties''. The prior odds for BayeScan was 10 for this analysis. We carried out BayeScEnv analyses using $p=0.5$ and $p=0$,
whereas $\pi$ was 0.1.

\begin{figure}
 \centering
 \includegraphics[width=\textwidth]{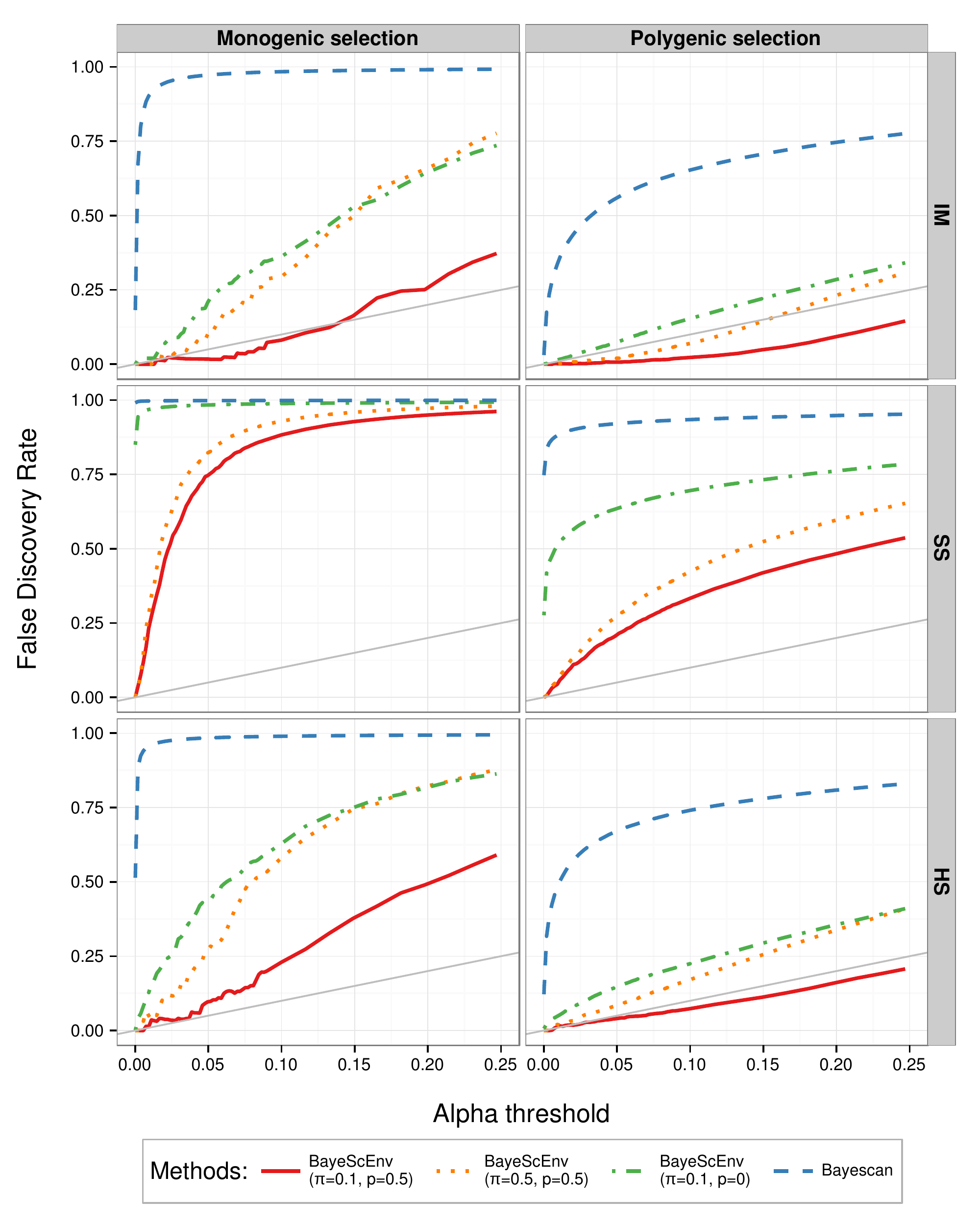}
 \caption{False Discovery Rate (FDR) against significance threshold $\alpha$ for three scenarios (IM: Island model,
 SS: Stepping-Stone model and HS: Hierarchically Structured model) and monogenic/polygenic selection. The grey line
 is the expected identity relationship between the FDR and $\alpha$. The models tested are BayeScan (blue dashed), 
 and BayeScEnv (orange dotted, green dot-dashed and solid red) with different probabilities $\pi$ of jumping away from 
 the neutral model (M1) and different preferences $p$ for the locus-specific model (M3). Note that $p=0$ means the 
 environmental model (M2) is tested against the neutral one only.}
 \label{fig_fdr}
\end{figure}

\begin{figure}
 \centering
 \includegraphics[width=\textwidth]{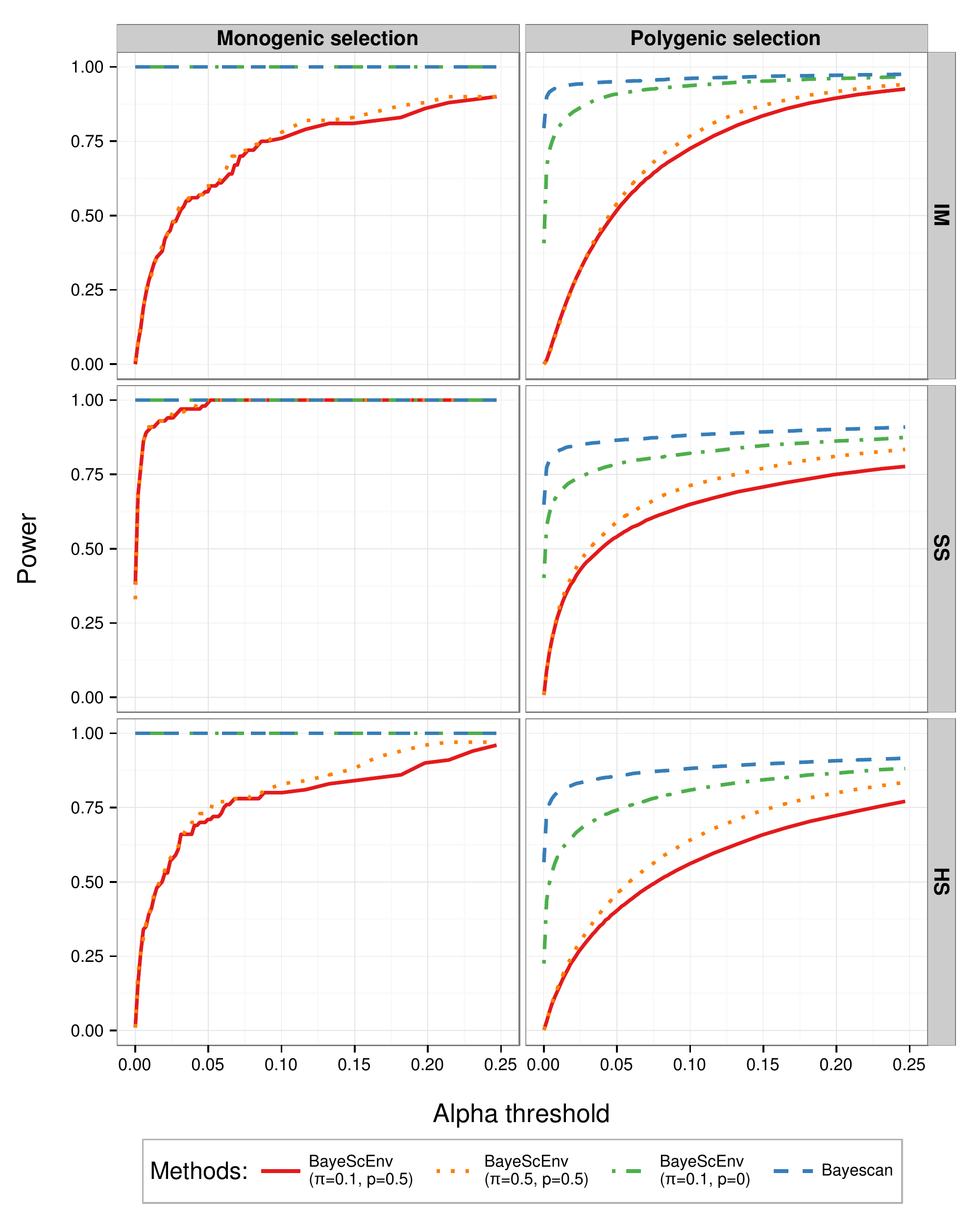}
 \caption{Power against significance threshold $\alpha$ for three scenarios (IM: Island model,
 SS: Stepping-Stone model and HS: Hierarchically Structured model) and monogenic/polygenic selection. The models tested 
 are BayeScan (blue dashed), and BayeScEnv (orange dotted, green dot-dashed and solid red) with different probabilities 
 $\pi$ of jumping away from the neutral model (M1) and different preferences $p$ for the locus-specific model (M3). Note that $p=0$ 
 means the environmental model (M2) is tested against the neutral one only.}
 \label{fig_pow}
\end{figure}

\begin{figure}
 \centering
 \includegraphics[width=\textwidth]{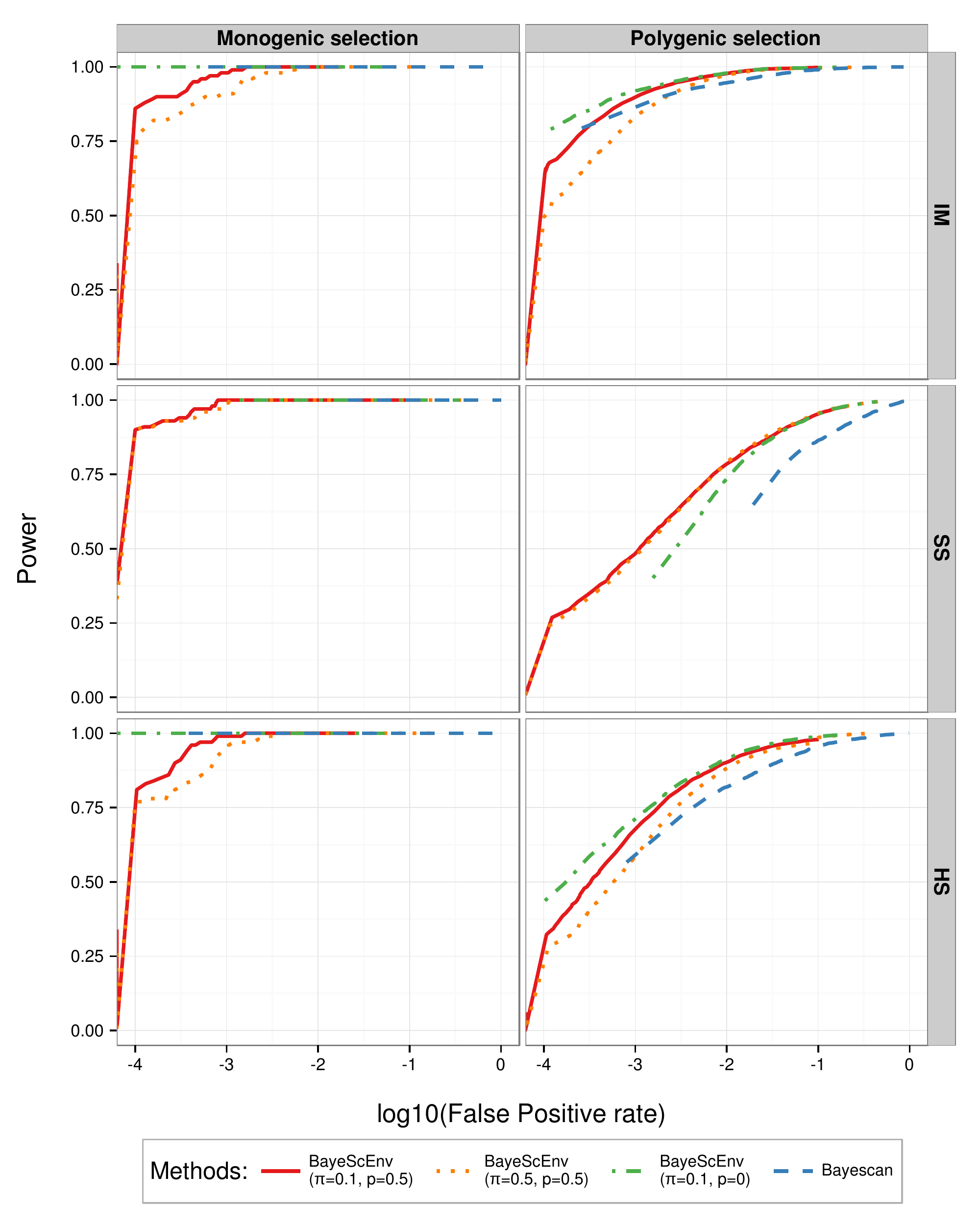}
 \caption{Power against False Positive Rate (FPR), a.k.a. ROC curve, for three scenarios (IM: Island model,
 SS: Stepping-Stone model and HS: Hierarchically Structured model) and monogenic/polygenic selection. The models tested 
 are BayeScan (blue dashed), and BayeScEnv (orange dotted, green dot-dashed and solid red) with different probabilities 
 $\pi$ of jumping away from the neutral model (M1) and different preferences $p$ for the locus-specific model (M3). Note that $p=0$ 
 means the environmental model (M2) is tested against the neutral one only.}
 \label{fig_ROC}
\end{figure}

\section*{Results}

\subsection*{Simulation results}

By definition, a threshold value of $\alpha$ used to decide whether $q$-values are significant or not is
expected to yield an FDR of $\alpha$ on the long run, when the model is robust and priors are calibrated.

As shown in Fig. \ref{fig_fdr}, BayeScan was less well calibrated, yielding higher FDRs than BayeScEnv under all
scenarios and for both monogenic and polygenic selection. Additionally, for BayeScEnv, the implementation using
$\pi=0.1$ was fairly well calibrated (i.e. the curve is close the grey line in Fig. \ref{fig_fdr}) under the IM 
scenario (for both monogenic and polygenic versions) and under the polygenic version of the HS scenario. This 
implementation was much more conservative than the one using $\pi=0.5$. For $\pi=0.1$ and $p=0$, the 
FDRs were closer to those yielded by BayeScan, but still lower.

The higher FDR for BayeScan and BayeScEnv with $\pi=0.5$ or $p=0$ was mainly driven by a higher 
FPR rather than a lack of power (Fig. \ref{fig_pow}, see also Fig. S3 in the SI). Notably though, BayeScan had a 
quite high power, higher than that of BayeScEnv. Note, however, that BayeScEnv with $p=0$ had, as BayeScan, a maximal 
power in the monogenic scenarios, and was almost as powerful as BayeScan in the polygenic scenarios. Yet its FDR was 
lower (sometimes much lower) than that of BayeScan. This indicates that the incorporation of environmental data helps to reduce 
the error rate both with or without the inclusion of spurious locus-specific effects ($\alpha_i$). More details 
regarding the FPR results are available in the Supplementary Information (Fig. S3). 

Another traditional way to apprehend the compromise between power and false positives is the so-called ROC curve,
plotting power against FPR (Fig. \ref{fig_ROC}). In these plots, the curve that is ``more to the left''
is preferred because this means it offers higher power for a lower FPR. Fig. \ref{fig_ROC} shows that BayeScEnv 
with $\pi=0.1$ and $p=0$ performed best under the IM and HS scenarios, whereas BayeScEnv with $\pi=0.1$ and
$p=0.5$ performed better under the ``harder'' SS scenario. Overall, although BayeScan has higher power to detect local adaptation, 
it is still too liberal when deciding that a locus is under selection for the scenarios we investigated.

\subsection*{Analysis of human data from Asia}

The results of the human dataset analysis (Table \ref{tab_hgdp}) show a dramatic discrepancy between the two methods. Whereas BayeScan yields a very large number
(66,316) of markers considered as significant at the 5\% threshold, many fewer markers (154 to 2728) are considered
significant by BayeScEnv. Gene Ontology (GO) enrichment tests identified many significant terms (Table \ref{tab_hgdp}). Note, however, that in the 
altitude and temperature analyses they correspond to a small number of genes (11 and 20 respectively, see Table \ref{tab_hgdp}). 
The number of genes is larger for the precipitation analysis (359) and even larger for the analysis using BayeScan (5628).

Regarding the altitude, significant biological processes included the fatty acid metabolism (e.g. SCARB1), skin pigmentation 
(e.g. MLANA, SLC24A5), kidney activity (e.g. SLC12A1) and oxido-reductase activity (e.g. NOS1AP
). Regarding the temperature, significant biological process included
cardiac muscle activity (e.g. SLC8A1) and development (e.g. NRG1, FOXP1), fatty acid metabolism (e.g. FADS1, FADS2) and response
to hypoxia (e.g. SLC8A1, SERPINA1). For the precipitation analysis with BayeScEnv, as well as the BayeScan analysis, the number of
significant terms was too large for hand-picked examples to be feasible.

\begin{figure}
 \centering
 \includegraphics[width=0.7\textwidth]{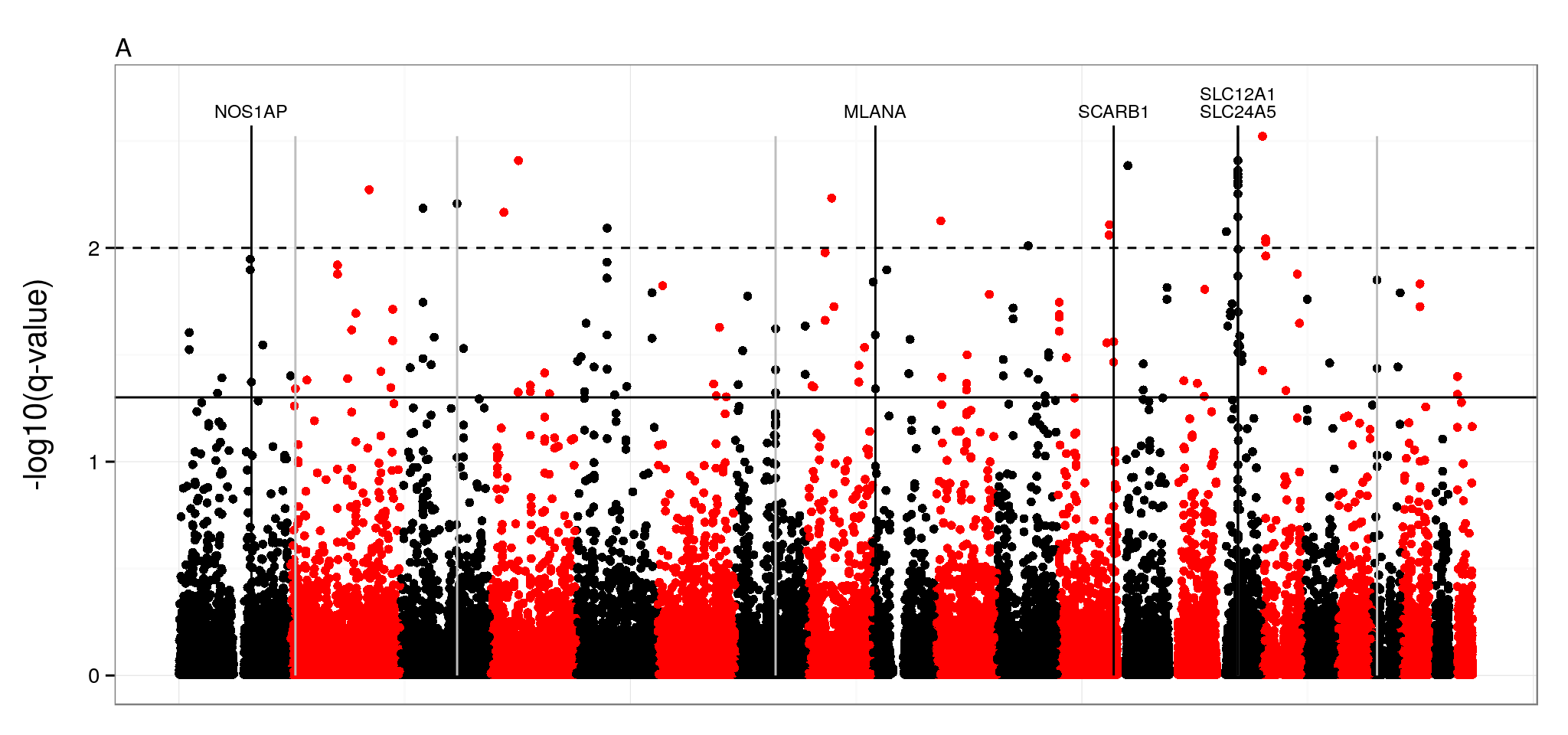}\\
 \includegraphics[width=0.7\textwidth]{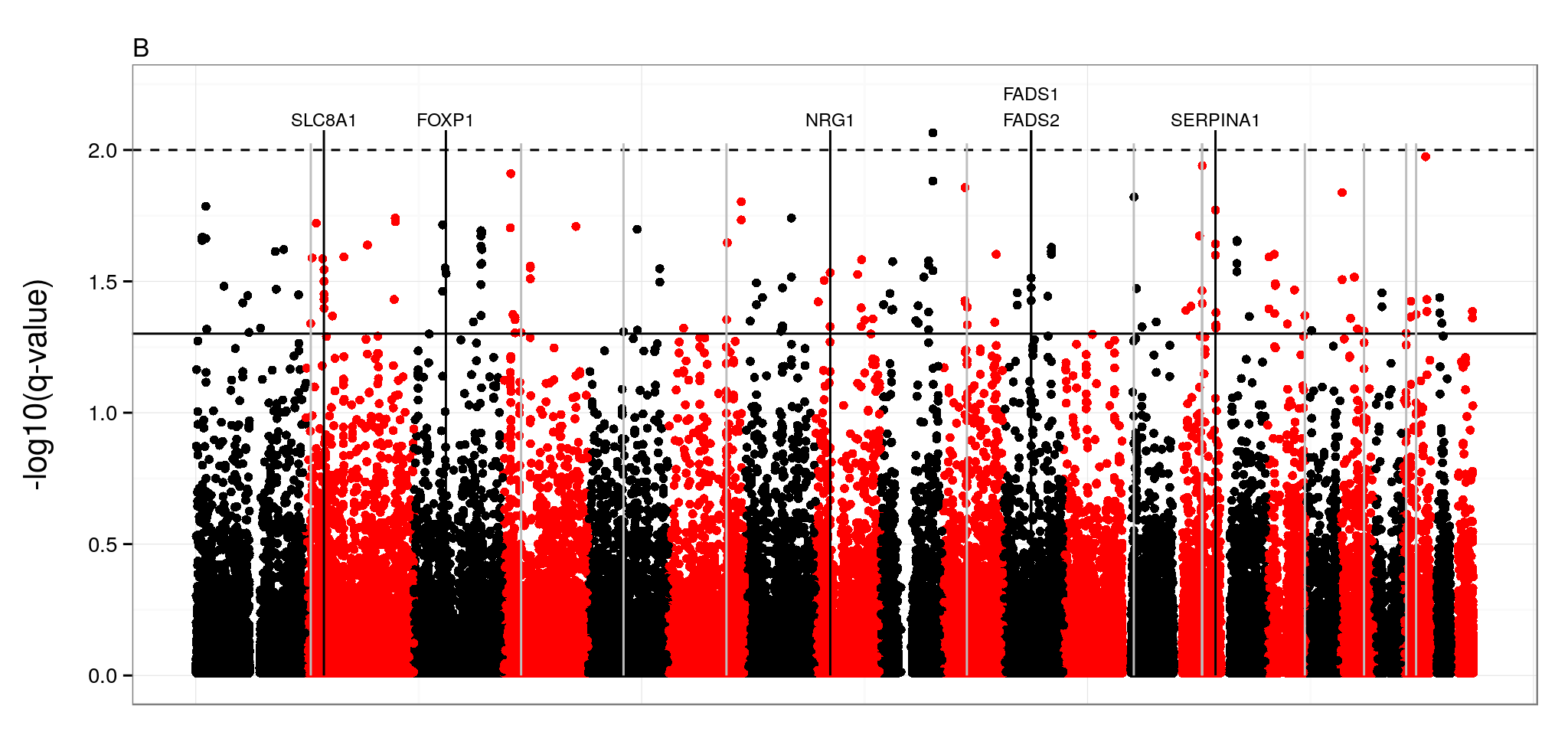}\\
 \includegraphics[width=0.7\textwidth]{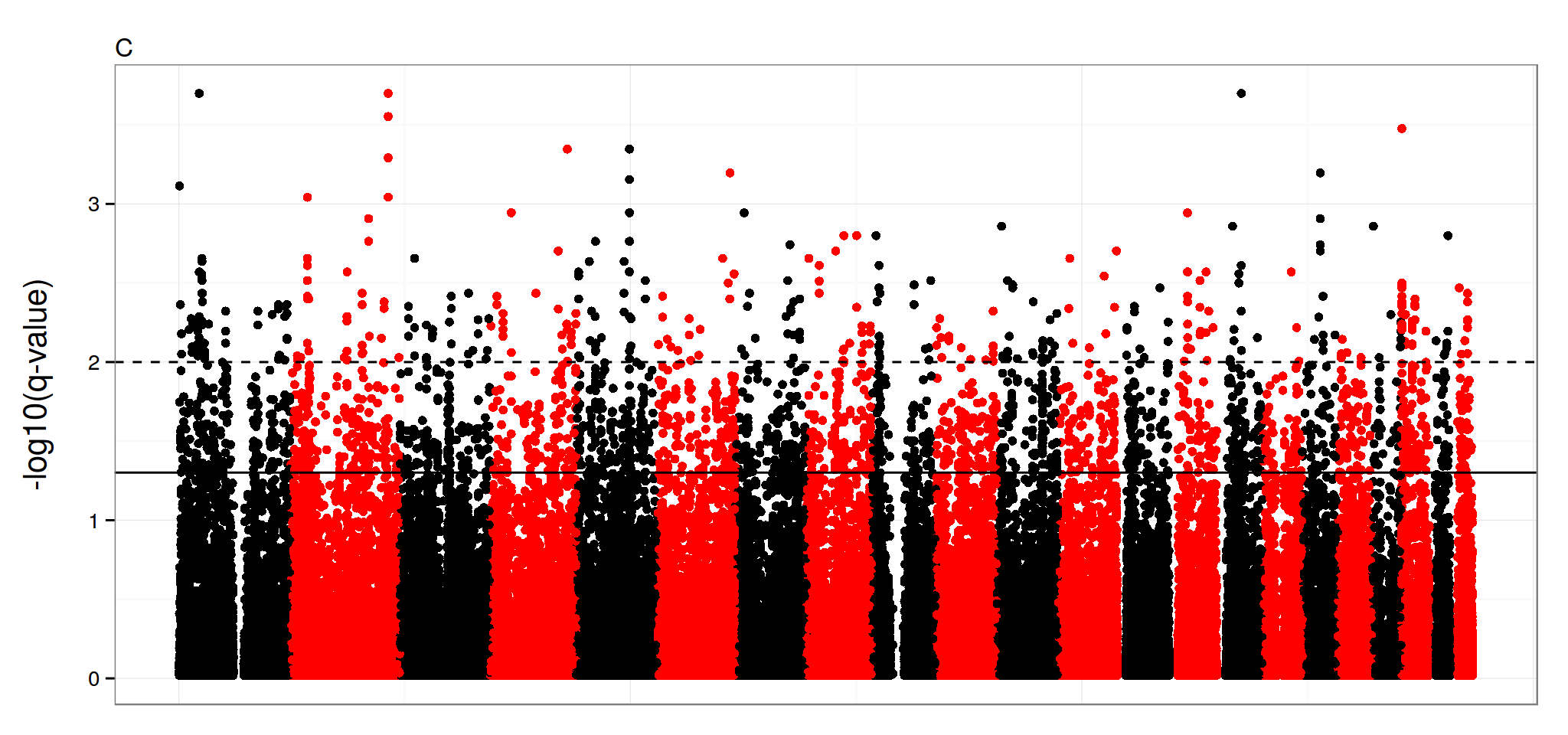}\\
 \includegraphics[width=0.7\textwidth]{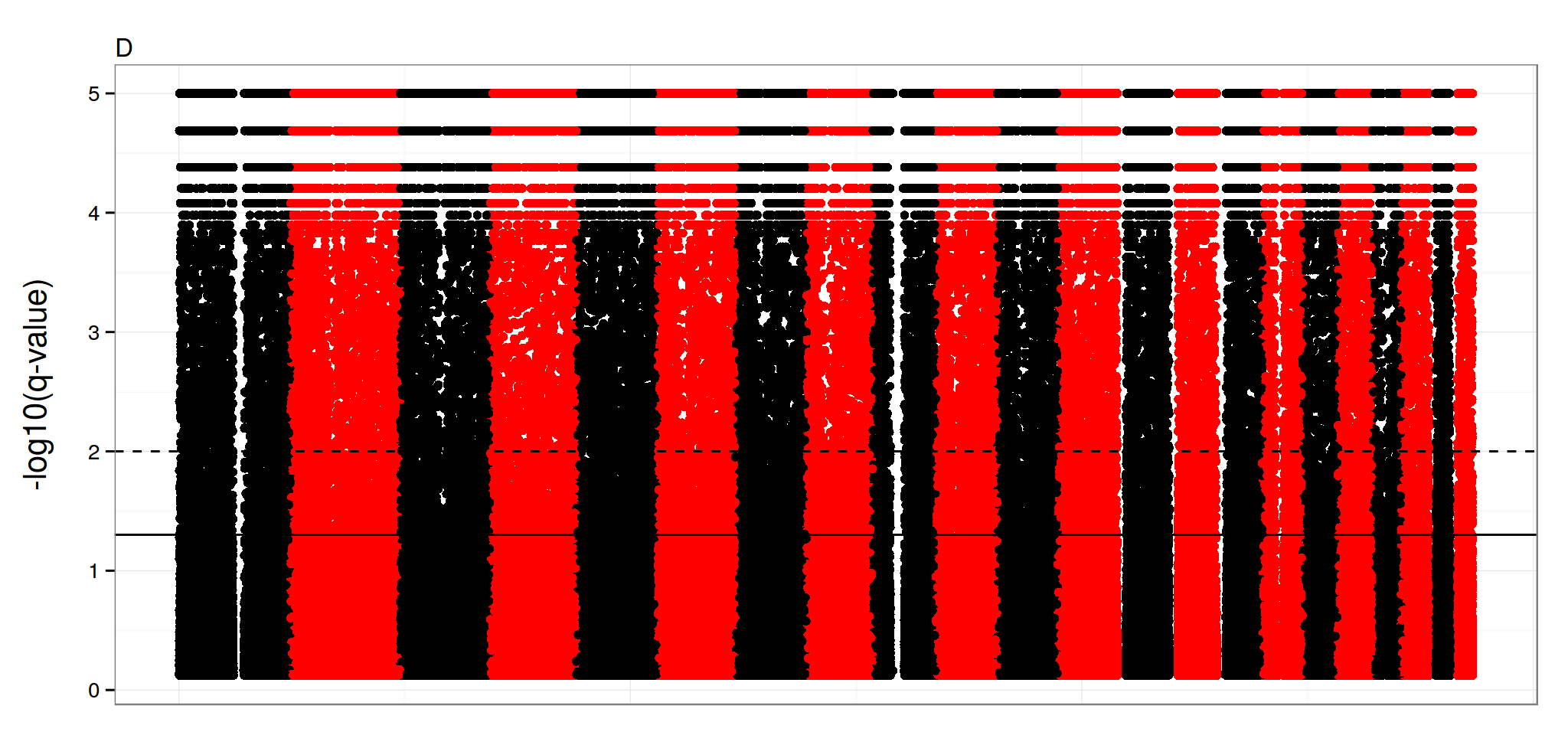}
 \caption{Manhattan plot of the $q$-values for the human dataset when using BayeScEnv with altitude (A), temperature (B),
 precipitations (C) or when using BayeScan (D). For altitude and temperature (A and B), genes mentioned in the text are
 displayed using black lines and genes associated with a significant GO term using grey lines. Top ``stripes'' for BayeScan
 (D) are artefacts due to finite number of iterations in RJMCMC (e.g. 0, 1, 2, 3... iterations outside of the non-neutral model), 
 corresponding to determined posterior probabilities when divided by the total number of iterations.}
 \label{fig_man}
\end{figure}

The significance results ($q$-values) are displayed as a Manhattan plot in Fig. \ref{fig_man}, along with the above mentioned
genes for the altitude and temperature analyses (Fig. \ref{fig_man}, A and B). Other regions of the genome also 
include outlier loci but they correspond to non-coding regions, or are close to genes associated to GO terms that were not significant, 
or to proteins without a known function (e.g. C9orf91, which was the most significant gene in the temperature analysis). Pattern of linkage disequilibrium
was visible, which sometimes strongly supported some candidate genes (Fig. \ref{fig_man}, A, SLC12A1 and SLC24A5). Finally, comparing
BayeScEnv (Fig. \ref{fig_man}, A, B and C) and BayeScan analyses (Fig. \ref{fig_man},D), we see that BayeScan yielded too many
significant markers for a Manhattan plot to be a useful display of the results. An interesting pattern is that BayeScan yielded far more outlier markers 
with maximal certainty (e.g. posterior probability of one) than BayeScEnv.
For the present dataset, 22,516 markers had a posterior probability of one, whereas the maximal posterior probability yielded by
BayeScEnv was 0.9998. Finally, almost all loci detected using BayeScEnv were also found when using BayeScan (between 98\% for altitude to
100\% for the two other variables).

\begin{table}
 \centering
 \begin{tabular}{cc|ccc}
 Method & Variable & \begin{minipage}{3cm}\centering Nr of significant SNPs\end{minipage} & \begin{minipage}{3cm}\centering Nr of significant GO terms\end{minipage} 
 & \begin{minipage}{4.1cm}\centering Nr of genes associated with a significant GO term\end{minipage}\\
 \hline
 \multirow{3}{*}{BayeScEnv} & altitude & 154 & 32 & 11\\
 & temperature & 170 & 103 & 20\\
 & precipitation & 2728 & 439 & 359 \\
 \hline
 BayeScan & --- & 66,316 & 469 & 5628\\
 \end{tabular}
 \caption{Results from BayeScan and BayeScEnv on the human dataset. FDR significance threshold was set to 5\%.
 The total number of tested markers was 446,117. }
 \label{tab_hgdp}
\end{table}

\subsection*{Analysis of Atlantic salmon data}

The results of the salmon dataset analyses (Table \ref{tab_salmon}) are again a clear demonstration of the discrepancy between the methods. Whereas BayeScan
yields 238 SNPs significant at the 5\% level, BayeScEnv yields between 0 and 8 significant markers with $p=0.5$ and between
45 and 62 with $p=0$. Thus, in agreement with the theoretical expectations and the simulation studies, we obtain more candidates loci 
with $p=0$. All loci obtained when using BayeScEnv (for any variable and any value of $p$) were found when using BayeScan. Unfortunately,
the Atlantic salmon genome is poorly annotated and, therefore, it was not possible to carry out a gene ontology enrichment analysis.

\begin{table}
\centering
\begin{tabular}{cc|ccc}
 Method & Variable & \multicolumn{2}{c}{Number of significant}\\
 & & with $p=0.5$ & with $p=0$\\
 \hline
\multirow{3}{*}{BayeScEnv} & temperature & 8 & 62\\
 & precipitations & 5 & 45 \\
 & river properties & 0 & 46\\
 \hline
 BayeScan & --- & \multicolumn{2}{c}{238}\\
 \end{tabular}
 \caption{Results from the BayeScan and BayeScEnv on the salmon dataset. FDR significance threshold was set to 5\%.
 The total number of tested markers was 2078.}
 \label{tab_salmon}
\end{table}

\section*{Discussion}

\subsection*{Features and performance of the method}

The method we introduce in this paper, BayeScEnv, has several desirable features. First, just as BayeScan, it is a model-based 
method. This means that the null model can be understood in terms of a process of neutral evolution. One can thus predict 
what the method is able to fit or not. Second, we explicitly model a process of local adaptation caused by an 
environmental variable. Third, in order to render the model more robust, we account for locus-specific effects unrelated to 
the environmental variable under consideration. These departures can be due to another process of local adaptation (i.e. caused by unknown 
environmental variables), to large differences in mutation rates across loci, to background selection 
\citep{charlesworth_background_2013} or complex spatial effects, such as allele surfing \citep{edmonds_mutations_2004}.
Our simulation results show that when compared to BayeScan, BayeScEnv has a better control of
its false discovery rate under various scenarios (Fig. \ref{fig_fdr}), yielding fewer, but more reliable 
candidate markers. Obviously, this has a cost in terms of absolute power (Fig. \ref{fig_pow}), but BayeScEnv still performs better 
than BayeScan in terms of the compromise between true and false positives (Fig. \ref{fig_ROC}).



Besides, the parametrisation of BayeScEnv allows for a fine and intuitive control of the false positive rate and power. For example,
setting $p$ to 0 increases both power and false positive rate, whereas setting $p=0.5$ will allow for a more conservative 
test. This is because with $p=0$, thus in the absence of the locus-specific effect model (M3), the local adaptation model (M2) will
absorb much of the signal in the data, yielding a higher probability of detecting true positives, but also a higher sensitivity to
false positives. Our simulation results show that, if the species under study has moderate to large dispersal abilities (c.f. hierarchical 
structure or island model), the former parametrisation will be more appropriate, whereas for species with low dispersal abilities (c.f. 
stepping-stone model) the latter should be preferred. Thus, being able to choose the right parametrisation only requires limited knowledge about the dispersal abilities of the species. 

We note that BayeScan was recently extended to consider species with hierarchical population structure 
\citep[BayeScan3,][]{foll_widespread_2014}. With BayeScan3 it is now possible to study widely distributed
species covering several continents or geographic regions. It is also possible to better focus on local adaptation by considering groups that include pairs of
populations inhabiting different environments such as low and high altitude habitats. Thus, BayeScan3, allows for the consideration of categorical 
environmental variables. Our new approach on the other hand, allows the study of local adaptation related to continuous environmental variables in 
species with a more restricted range.


\subsection*{How to quantify `environmental differentiation'?}

To model local adaptation, we compute an ``environmental differentiation'' in terms of the Manhattan distance to a reference value. 
Although this reference can conveniently be chosen as the average of the 
environmental values across the sampled populations, other kinds of reference may be biologically more relevant.
For example, in our analysis of the effect of elevation in humans, it seems appropriate to use sea level as the reference. 
Indeed, given the kind of environmental variables elevation is a proxy for (e.g. partial pressure of oxygen, 
temperature, solar radiation, etc.), for most systems we would consider the sea level as a neutral environment rather than the 
differentiated one. 

Another way to account for environmental differentiation is to use Principal Component Analysis (PCA), 
providing one of the axes to BayeScEnv as a description of the distance between environments. Despite this practice being an elegant way to
summarise environmental distance between populations, it also has the drawback of making it more difficult to identify the ``causal'' variable. 

Note that the environmental variables must be standardised so as to avoid scale inconsistencies between $g$ and $\alpha$ and $\beta$. 
If we choose the average environmental value as reference, then standardisation 
involves mean-centring and rescaling to have unit variance. However, if we choose another reference, then standardisation only 
involves rescaling to have unit variance.

\subsection*{Comparison with other environmental association methods}

There are several genome-scan approaches that incorporate environmental information. Some are mechanistic 
(e.g. Bayenv, \citealp{coop_using_2010}) while others are phenomenological 
(e.g. LFMM, \citealp{frichot_testing_2013} and gINLAnd, \citealp{guillot_detecting_2014}). These methods 
perform a regression between allele frequencies and environmental values.
Yet non-equilibrium situations combined with complex spatial structuring can lead to spatial correlations in allele frequencies, 
which in turn can lead to high false positive rates. To minimise this problem, the above methods take into account allele
frequency correlations across populations while performing the regression.


BayeScEnv, on the other hand, assumes that all populations
are independent, exchanging genes only through the migrant pool. However, it includes 
a locus-specific effect unrelated to the environmental variable that helps to take into account locus-specific spatial effects due to deviations from the 
underlying demographic model. The fact that this approach works is illustrated by our simulation study, which showed that BayeScEnv was 
fairly robust to isolation-by-distance and a hierarchically structured scenario. Moreover, the analyses of simulated datasets from \citet{de_villemereuil_genome_2014}, available in the SI, show that even under very complex scenarios, BayeScEnv can compete with other 
environmental association methods. Nevertheless, we note that BayeScEnv is best suited for species with medium to high dispersal abilities 
such as marine species and anemophilous plants. 


Another point that distinguishes BayeScEnv from these methods is that it does not assume any particular functional 
form for the relationship between environmental values and allele frequencies. While existing association methods all assume a 
clinal pattern, BayeScEnv only assumes that genetic differentiation increase exponentially with environmental differentiation. 
This allows for a more diverse family of relationships between allele frequencies and the environment. For example, a scenario where the
same allele is favoured at the margins and counter selected in the middle of the species range can be studied with BayeScEnv but 
would certainly represent a problem for the other association methods. Such a scenario would arise, for instance, when the target of selection
in extreme environments are plasticity genes \citep{morris_gene_2014}, or genes regulating stress response. 
Alternatively, the chosen environmental variable might actually be a proxy for another selective variable that takes similar values when the former takes very low or very high values. For example, 
environments with very low or very high temperatures are often also arid environments. Another difficult scenario that BayeScEnv 
would be able to detect is one in which two populations with very similar values for the environmental factor have very different alleles frequencies at a locus and both experience environmental conditions very different from those of the other populations. Such patterns are difficult to relate to local adaptation and might most likely 
be caused by high drift in extreme environments, due to reduced population sizes. Yet, such a signal should in principle be  
captured by the $\beta$ parameter. One particular selective scenario that could explain such a pattern would be one of positive
frequency-dependent selection modulated by the environment (i.e. only extreme environments would induce selection),
as expected in the case of Mullerian mimicry with differential predator pressure \citep{borer_positive_2010}.
Nevertheless, the number of species where such a scenario would be biologically plausible is limited. In any case, loci showing 
such a pattern can be easily identified by \textit{post-hoc} inspection of their allele frequencies in the different populations. It would 
then be possible to label these loci as false positives if frequency-dependent selection is deemed an unlikely scenario. All
these scenarios are illustrated in the SI.

Finally, BayeScEnv is one of the very few methods to study gene-environment associations that can be used with dominant data
\citep[but see also][]{guillot_detecting_2014}.

\subsection*{Data analysis}

When confronted with real datasets, BayeScEnv typically returned fewer significant markers than BayeScan. This is explained both
by the focus on searching for outliers linked to a specific environmental factor and by the lower false positive rate of our approach. 
When applied to the human dataset, BayeScEnv identified several genomic regions that are enriched
for gene ontology terms relevant to potential local adaptation to altitude or 
temperature. We emphasise that this study was not meant to exhaustively and rigorously investigate local adaptation in Asian
human populations. However, our results tend to demonstrate that the candidates yielded by BayeScEnv have a biological interpretation.
For example, skin pigmentation and cardiac
activity could clearly be involved in responses to increased solar radiation and depleted oxygen availability at high elevation.


Much of the ontologies linked to temperature were potentially confounded with adaptation to altitude, such as the response
to hypoxia and cardiac muscle activity. Also, fatty acid metabolism was associated to both altitude and temperature.
Of course, the biological functions described here do not account for all the signals yielded by BayeScEnv
(see Fig. \ref{fig_man}, A and B).
Other genomic significant regions include genes with less obvious biological function regarding local 
adaptation, non-coding regions and proteins without a known function.
Finally, the analysis using the precipitation variable yielded too many significant markers for
a detailed analysis of the biological functions involved. This may not necessarily be due to a confounding effect of the spatial structure
(the human Asian populations being structured mainly from West to East, while the Eastern climate is characterised by strong precipitations during the monsoon), since precipitation may behave as a surrogate for several environmental variables.

As the Atlantic salmon genome is poorly annotated, we could not identify genes associated to the observed outlier loci. 
However, the discrepancy between the number of candidates yielded
by BayeScan and BayeScEnv was still quite impressive in this case. Also, when using the parametrization $p=0$, we obtained almost an order of magnitude
more candidates (though our simulations tend to demonstrate that this was probably at a cost of a larger false positive rate). 

\subsection*{Conclusion}
The main improvement introduced by our new method, BayeScEnv, over existing $F_{ST}$-based genome-scan approaches is the possibility of focusing 
on the detection of outlier loci linked to genomic regions involved in local adaptation and better distinguishing between the signal of positive 
selection and that of other locus-specific processes such as mutation and background selection. Although it does not explicitly model complex
spatial effects, the consideration of two different locus-specific effects make it more robust to potential deviations from the migrant pool model. 
This is reflected in its much lower false discovery rate when compared to BayeScan. 

Our new formulation also allows for an improved control of the true/false 
positives compromise through the parameter $p$, which describes our preference for the model that includes a locus-specific effect unrelated to the 
environmental factor over the model that includes environmental effects. Although we recommend using $p=0.5$, lower values (including 0) could be used 
if population structure is weak or maximising power is more important than reducing the false positive rate. 

With this new method, there are now three alternative formulations of genome-scan methods based on the $F$ model. BayeScan detects a wide range of 
locus-specific effects (including background selection). Although its false discovery rate is higher than that of the two extensions, it is able to detect regions 
of the genome subject to purifying selection. The hierarchical version of this original formulation, BayeScan3, allows the study of local adaptation due to categorical 
environmental factors. Finally, our new method, BayeScEnv, is more appropriate to detect genomic regions under the influence of selective pressures exerted by 
continuous environmental variables. Thus, all three methods are complementary and jointly cover scenarios applicable to a wide range of species

\section*{Acknowledgement}
We thank M. Foll for providing the source code of BayeScan and for clarifying several issues related to the code, J. Renaud for 
his help on getting the average altitude out of the HGDP latitude/longitude data, S. Schoville for the
BIOCLIM data, E. Bazin for his help on the HGDP data analysis and V. Bourret for his help on the salmon dataset. 
PdV was supported by a doctoral studentship from the French \textit{Minist\`{e}re de la Recherche et de 
l'Enseignement Sup\'{e}rieur}. OEG was supported by the Marine Alliance for Science and Technology for Scotland (MASTS). 

\bibliography{biblio}
\bibliographystyle{molecol}

\section*{Data Accessibility}
The Python code used to simulate data is available online in the Supplementary Information. The software is available online
at GitHub: \url{http://github.com/devillemereuil/bayescenv}.

\section*{Author contributions}
PdV and OEG designed the statistical model. PdV modified the C++ code and performed the simulation and data analysis. PdV and OEG
wrote the article.

\newpage

\appendix
\setcounter{figure}{0}
\setcounter{table}{0}
\setcounter{equation}{0}
\renewcommand{\thefigure}{S\arabic{figure}}
\renewcommand{\thetable}{S\arabic{table}}
\renewcommand{\thesubsection}{\arabic{subsection}}

\section*{Supplementary Information}

\subsection{Definition of the prior probabilities of jump between models}

Recall the three models of which we want to infer posterior probabilities:
\begin{description}
 \item[M1] Neutral model: $\log(\frac{1}{\theta_{ij}}) = \beta_{j}$,
 \item[M2] Local adaptation model with environmental differentiation $\log(\frac{1}{\theta_{ij}}) = \beta_{j} + g_{i}E_{j}$,
 \item[M3] Locus-specific model: $\log(\frac{1}{\theta_{ij}}) = \alpha_{i}+\beta_{j}$.
\end{description}
Let $\Pi_{2}$ be the prior probability of model M2 and $\Pi_{3}$ the prior probability of model M3. We assume that
the probability of going from M1 or M2 to the model M2 is equal to $\Pi_{2}$ (the same reasoning applies for M3 and $\Pi_{3}$), 
which leads to the following transition matrix:
\begin{equation}
\left(\begin{array}{ccc}
1-\Pi_{2}-\Pi_{3} & \Pi_{2} & \Pi_{3}\\
(1-\Pi_{2})(1-\Pi_{3}) & \Pi_{2} &  \Pi_{3}(1-\Pi_{2})\\
(1-\Pi_{2})(1-\Pi_{3}) & \Pi_{2}(1-\Pi_{3}) & \Pi_{3} 
 \end{array}\right).
\end{equation}
If we consider $\Pi_{2}=\pi (1-p)$ and $\Pi_{3}=\pi p$ where $\pi$ is the probability of jumping away from model M1 and $p$
the ``preference'' for model M3 (i.e. the probability of choosing the model M3 instead of model M2, when jumping away from model M1),
then we can write:
\begin{equation}
\left(\begin{array}{ccc}
1-\pi  & \pi (1-p) & \pi p\\
1-\pi+\pi^{2}p(1-p) &  \pi (1-p) & \pi p-\pi^{2}p(1-p) \\
1-\pi+\pi^{2}p(1-p) & \pi (1-p)-\pi^{2}p(1-p) & \pi p
 \end{array}\right).
\end{equation}
Thus, when $\pi$ is small (in practice $\pi<0.5$), the transition between models depends only very slightly on the current state of
the model, and the prior probabilities of each model reduce approximately to (ignoring second order terms)
\begin{equation}
 \begin{array}{rcl}
 P(\text{\textbf{M1}}) & = & 1-\pi, \\
 P(\text{\textbf{M2}}) & = & \pi (1-p), \\
 P(\text{\textbf{M3}}) & = & \pi p.\\
 \end{array}
\end{equation}

\subsection{Reversible jumps between the models}

According to \citet{brooks_markov_1998} and \citet{gelman_bayesian_2004}, the jump from model $l$ to model $k$ should be accepted
with a probability $\text{min}(r,1)$, with
\begin{equation}\label{eq_SI_jump}
 r = \frac{L(Y|\theta_{k},M_{k})P(\theta_{k}|M_{k})P(M_{k})}{L(Y|\theta_{l},M_{l})P(\theta_{l}|M_{l})P(M_{l})}
 \frac{J_{k\rightarrow l} J(u_{k}|\theta_{k},k,l)}{J_{l\rightarrow k} J(u_{l}|\theta_{l},l,k)}
 \left|\frac{\nabla g_{l,k}(\theta_{l},u)}{\nabla(\theta_{l},u)} \right|,
\end{equation}
where $\theta_{\bullet}$ is the parameter vector for model $\bullet$, $L$ stands for the likelihood of parameters $\theta_{\bullet}$ assuming
the model M${\bullet}$, and $J$ is the proposal kernel for the (potentially) new parameter $u_{\bullet}$. Note also the presence of 
the proposal of new model $J_{\circ\rightarrow\bullet}$.

Because a jump toward one of the two alternative models is proposed at each iterations, $\frac{J_{k\rightarrow l}}{J_{l\rightarrow k}}$
simplifies to one. Likewise, since the transformations from $\theta_{l}$ to $\theta_{k}$ only consists in setting some parameters to
0, the Jacobian determinant $\left|\frac{\nabla g_{l,k}(\theta_{l},u)}{\nabla(\theta_{l},u)} \right|$ also simplifies to one.

The most efficient way to propose a value for $u_{\bullet}$ is to draw from its own posterior \citep{brooks_efficient_2003}.
Thus, pilot runs are carried out before the actual reversible jump MCMC in order to approximate the posteriors of $\alpha_{i}$ and 
$g_{i}$ for each locus $i$. During these runs, the parameters $\beta_j$ are inferred alone (M1 model), whereas the parameters 
$\alpha_i$ and $g_i$ are inferred using models containing $\beta_j$ (M2 and M3 models). The posterior mean and variance obtained 
for these parameters from the pilot runs are used to parametrise Normal distributions. These distributions are used to propose a 
new value of a parameter when a jump to a model including it is proposed. Note that, since $g_{i}$ cannot be negative, a truncated 
Normal is used, and the $J$ kernel is modified accordingly in Eq. \ref{eq_SI_jump}.

\subsection{Statistical tests}
Using the posterior probability of model \textbf{M2} for the locus $i$, $P_{i}(\text{\textbf{M2}}|\bm{a}_{i},E)$, we can
calculate the posterior error probability \citep[PEP,][]{kall_posterior_2008} for locus $i$ as
\begin{equation}\label{eq_pep}
 \text{PEP}_{i} = 1 - P_{i}(\text{\textbf{M2}}|\bm{a}_{i},E).
\end{equation}
In order to calculate the $q$-value \citep{storey_positive_2003,muller_fdr_2006}, 
we rank the PEP$_{i}$ from from the lowest to the highest value, and define the 
$q$-value for locus $i$ as the average PEP for all loci having a PEP lower, or equal to, PEP$_{i}$:
\begin{equation}\label{eq_qval}
 q_{i} = \frac{1}{i}\sum_{k,k\leq i} \text{PEP}_{k}.
\end{equation}
Note that, because we calculate the average using only PEPs that are lower than PEP$_{i}$, we have 
$q_{i}\leq\text{PEP}_{i}$ for all $i$. The equality only holds for the minimal PEP(s). \Citet{kall_posterior_2008} 
advocate the use of the $q$-value because it is optimal in the sense of Bayesian classification theory 
\citep[see also][]{storey_positive_2003}. Our code outputs both PEPs
and $q$-values.\\
Both of these test statistics are strongly related to the control 
of False Discovery Rate \citep[FDR,][]{storey_direct_2002} during multiple testing. Contrary to the commonly used 
False Positive Rate (FPR), which is the probability of declaring a locus as positive given that it is actually 
neutral, the FDR is the proportion of the positive results that are in fact false positives. 
Note that the PEP is a ``locally'' (i.e. regarding only the focal locus) inferred measure of the 
FDR \citep[see][]{kall_posterior_2008}, whereas the $q$-value is based on inferring what the FDR would be when stating 
that the focal locus, and all the loci with a lower score should be considered as positives. \\
In the following, we will focus on the $q$-value of the local adaptation model (M2). Since we have a strong 
uncertainty regarding the biological origin of the locus-specific effect $\alpha_i$, we can consider it as a ``nuisance'' parameter 
in this particular inference framework.

\subsection{Hierarchically Structured (HS) scenario}

\begin{figure}[H]
 \centering
 \includegraphics[width=0.49\textwidth]{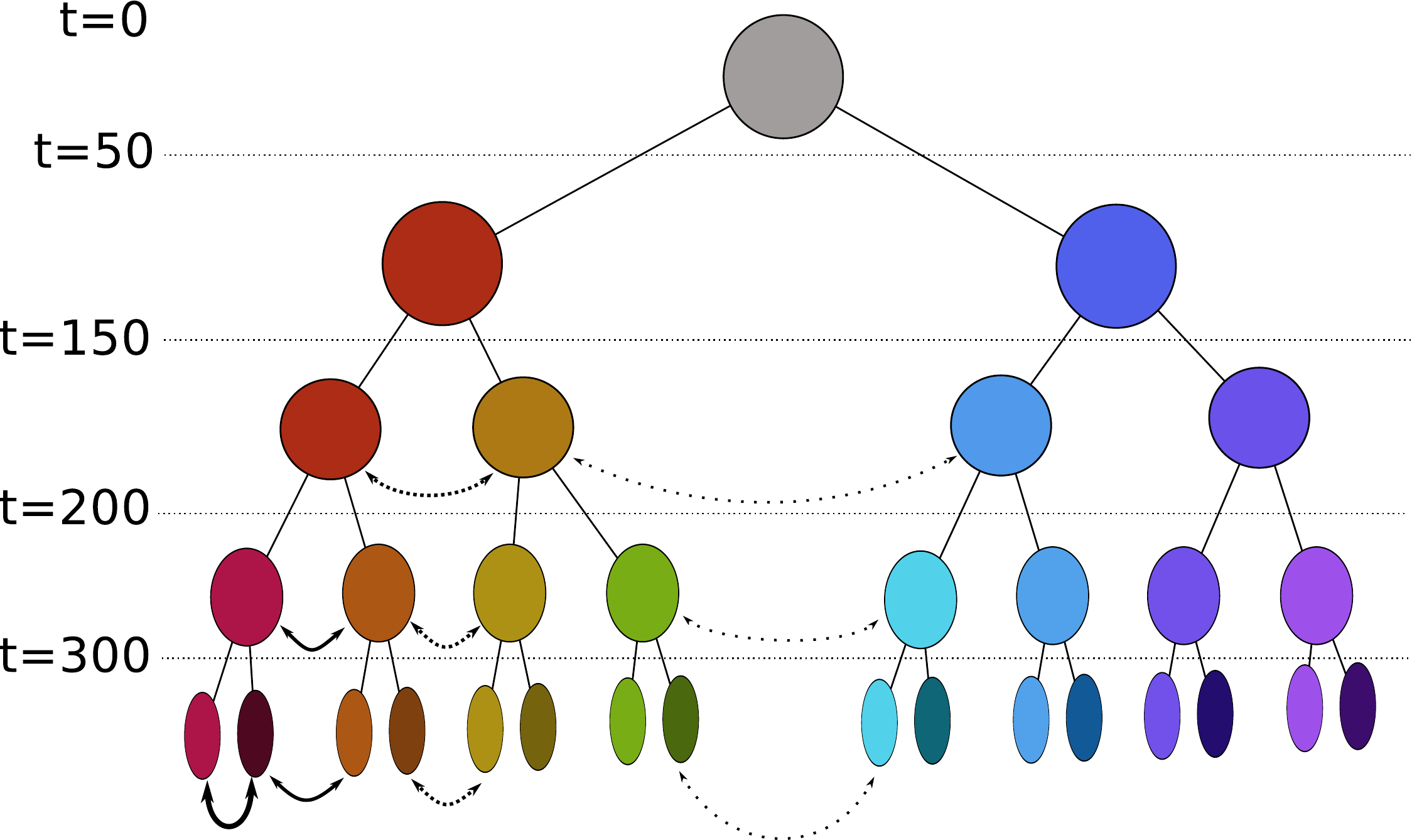}\\
 \caption{Schematic representation of the Hierarchically Structured scenario (HS). Fission events are shown as
 connectors and migration is denoted using double-arrow (thickness illustrate the strength of migration).}
 \label{suppfig_demo}
\end{figure}

Fig. \ref{suppfig_demo} gives a schematic representation of the demographic
scenario called HS in the main text, showing the fission events and the migration 
between populations. Note that only some illustrative migration combinations are represented 
for the sake of simplicity. An important feature of this model is that the probability of migration between two
populations decreases as their relatedness (measured by the number of fission events separating them) decreases. A full model 
description can be found in \citet{de_villemereuil_genome_2014}.

\subsection{Simulation scripts}

Four Python scripts are provided as Supplementary Files:
\begin{description}
 \item[\texttt{IM\_mono.py}] Island model with monogenic selection
 \item[\texttt{IM\_poly.py}] Island model with polygenic selection
 \item[\texttt{SS\_poly.py}] Stepping-stone model with polygenic selection
 \item[\texttt{HS\_poly.py}] Hierarchically structured model with polygenic selection
\end{description}
These scripts require Python 2.7 and SimuPOP 1.1. Note that the monogenic version is provided only for the 
island model, as the modification are identical for the two other models.

\subsection{Environmental gradient}

\begin{figure}[H]
 \centering
 \includegraphics[width=0.49\textwidth]{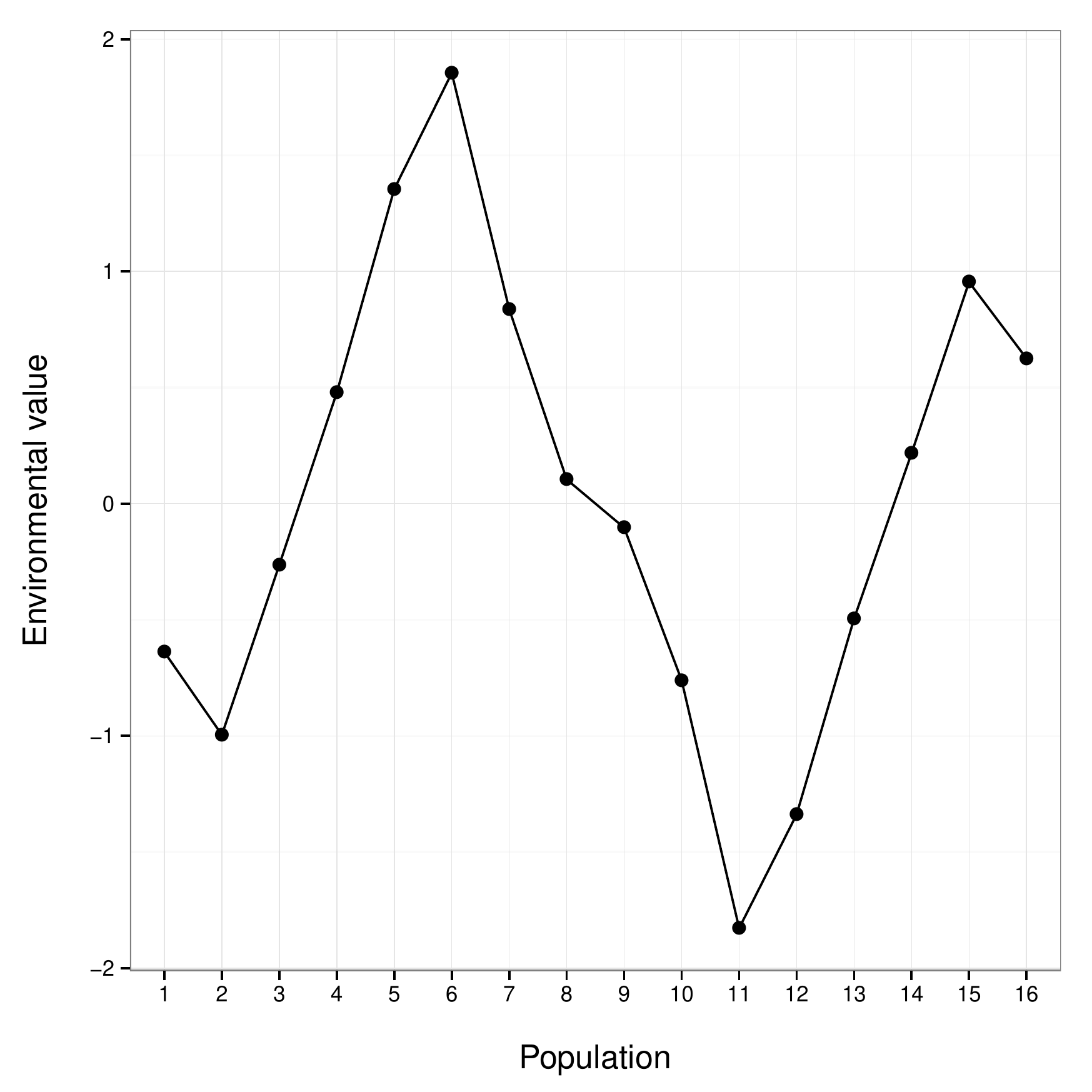}
 \caption{Environmental value for all 16 populations in the scenarios SS and HS.}
 \label{suppfig_env}
\end{figure}

The environmental gradient was designed to be independent from (i.e. not be confounded with) the 
population structure for the scenarios SS and HS (see Fig. \ref{suppfig_env}).

This environmental variable was already standardised for its use in BayeScEnv.
A transformation of this variable was used to define selective pressure on an appropriate scale:
\begin{equation}
 s_{j} = s_{0}\frac{1-e^{-E_{j}}}{1+e^{-E_{j}}},
\end{equation}
where $s_{0}$, the strength of the selection was chosen to be 0.1 for the monogenic case and 0.02 for the polygenic
case. The individual fitness was calculated in a multiplicative fashion:
\begin{equation}
 W = (1+s_{j})^{n_{11}}(1-s_{j})^{n_{00}},
\end{equation}
where $n_{11}$ and $n_{00}$ are the number of loci at which the individual is homozygous for the advantageous and disadvantageous 
allele, respectively. Note that this fitness function assumes co-dominance, with a heterozygous fitness of 1.

The recombination rate was set to 
0.002 (one recombination between two adjacent loci, \textit{per} population and \textit{per} generation).
The mutation rate was set to $10^{-7}$ \textit{per} generation at every locus. The allele frequencies were initialised
using a beta-binomial distribution truncated between 0.1 and 0.9 to avoid too many monomorphic loci.

\subsection{Simulation results for the False Positive Rate}

\begin{figure}[H]
 \centering
 \includegraphics[width=0.85\textwidth]{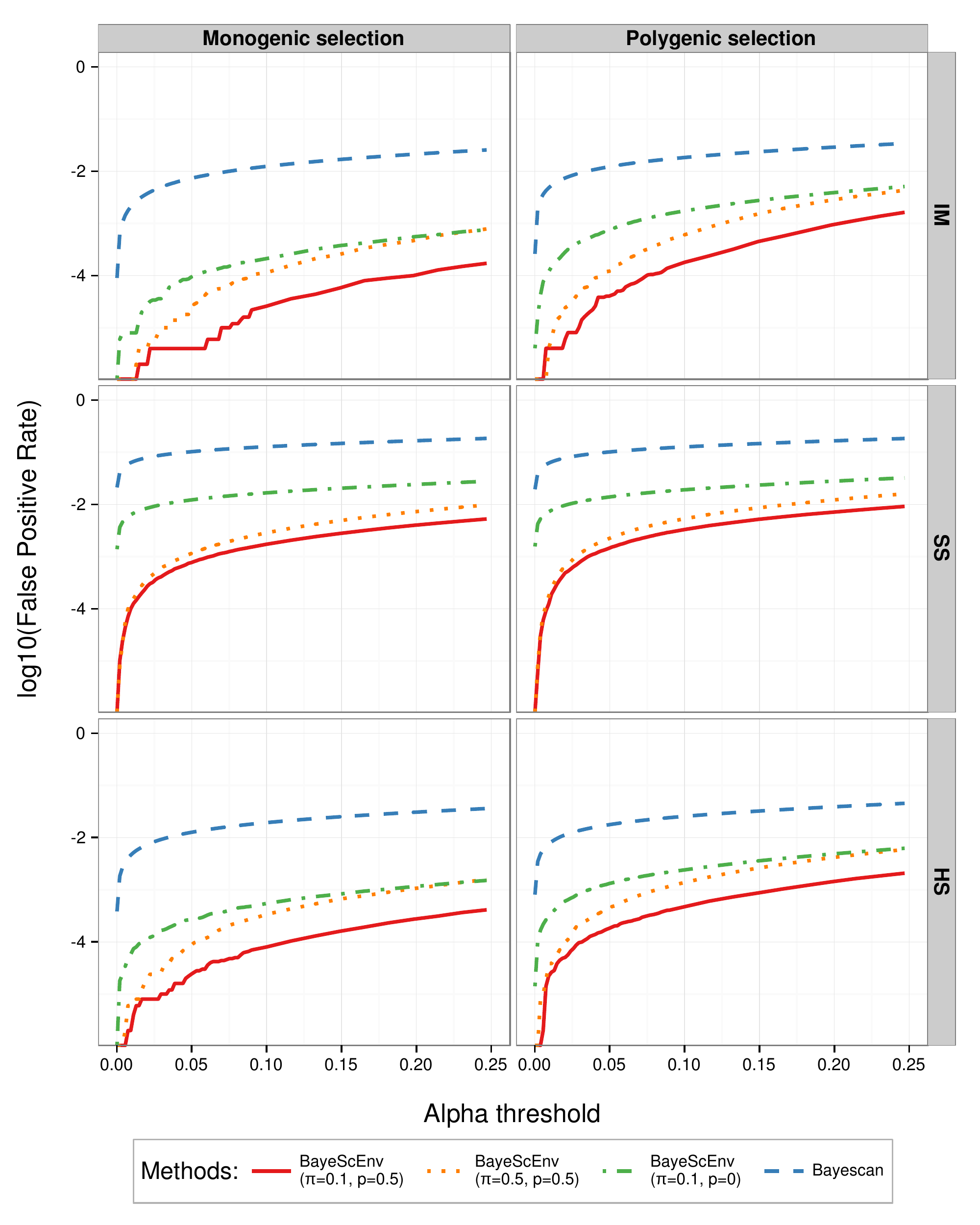}
 \caption{False Positive Rate (FPR) against significance threshold $\alpha$ for three scenarios (IM: Island model,
 SS: Stepping-Stone model and HS: Hierarchically Structured model) and monogenic/polygenic selection. The models 
 tested are BayeScan (blue dashed), and BayeScEnv (orange dotted, green dot-dashed and solid red) with different 
 probabilities $\pi$ of jumping away from the neutral model and different preferences $p$ for the locus-specific 
 model. Note that $p=0$ means the environmental model is tested against the neutral one only.}
 \label{supp_fig_fpr}
\end{figure}

The results regarding the False Positive Rate (FPR, Fig. \ref{supp_fig_fpr}) were qualitatively comparable to the
results regarding the False Discovery Rate (FDR, Fig. 2, main text). Indeed, we again found that the
most error-prone method was BayeScan, where BayeScEnv yielded fewer false positives. For the latter, the parametrisation 
$\pi=0.1$, $p=0.5$ was, as expected, the most conservative, whereas $\pi=0.1$, $p=0$ was the most laxist.

\subsection{List of candidate genes associated with significant GO terms}

Below is a list of the genes that fulfill the following two criteria. First, there is at least one significant SNPs in their neighbourhood,
indicating them as potential candidates for local adaptation. Second, at least one of their associated GO terms were found to be significantly 
enriched for candidates compared to the rest of the genome.
\begin{itemize}
 \item \textbf{For the altitude analysis:} SCARB1, SLC12A1, MUCL1, DNM2, MLANA, ATP6V1C2, CLDN12, FBN1, OTUD7A, SLC24A5, NOS1AP, SLC12A8
 \item \textbf{For the temperature analysis:} SYNE2, SPTB, ANKRD46, HAO1, HCK, FOXP1, ONECUT2, CDH15, ATP8A2, FADS2, ESR2, ATP6V1C2, FADS1, NRG1, APBB2, CMYA5, SERPINA6, SLC8A1, PRKG1, LAMA2, SERPINA1
\end{itemize}
Note that the majority of the significant GO terms were represented by only one gene for the altitude and temperature
analysis. The list of genes fulfilling the two criteria is not shown for the precipitation analysis and BayeScan, as there are
too many of them.

\subsection{Analysis of the simulation scenarios from \citet{de_villemereuil_genome_2014}}

\paragraph{Scenarios}
Given the computationally expensive nature of the simulations necessary to generate synthetic data, we compare BayeScEnv with methods other than BayeScan 
using the simulated datasets from \citet{de_villemereuil_genome_2014}. For more information regarding these scenarios, 
please refer to the article. Briefly, four polygenic scenarios were tested:
\begin{description}
 \item[HsIMM-C] Hierarchical scenario with a clinal environment following population structure
 \item[HsIMM-U] Hierarchical scenario with a random environment strongly correlated with population structure
 \item[IMM] Isolation with Migration Model
 \item[SS] Stepping-Stone model with a clinal environment, following the clinal population structure
\end{description}

\paragraph{Foreword}
These scenarios were tested against BayeScan \citep{foll_genome-scan_2008}, Bayenv \citep{coop_using_2010} and LFMM 
\citep{frichot_testing_2013}. Note that these scenarios are very difficult for all methods, thus we do not expect the 
FDR to be well calibrated. Also, in contrast with the study in the main text, \citeauthor{de_villemereuil_genome_2014} used a prior odds of 
100 instead of 10 for BayeScan.\\
When interpreting the results, it should be remembered that the FDR depends on both the FPR
and power. All things being equal, the FDR will be higher if the FPR is higher, and lower if the power is higher.

\begin{figure}[H]
 \begin{minipage}{0.5\textwidth}
 \begin{center}
     \includegraphics[width=\textwidth]{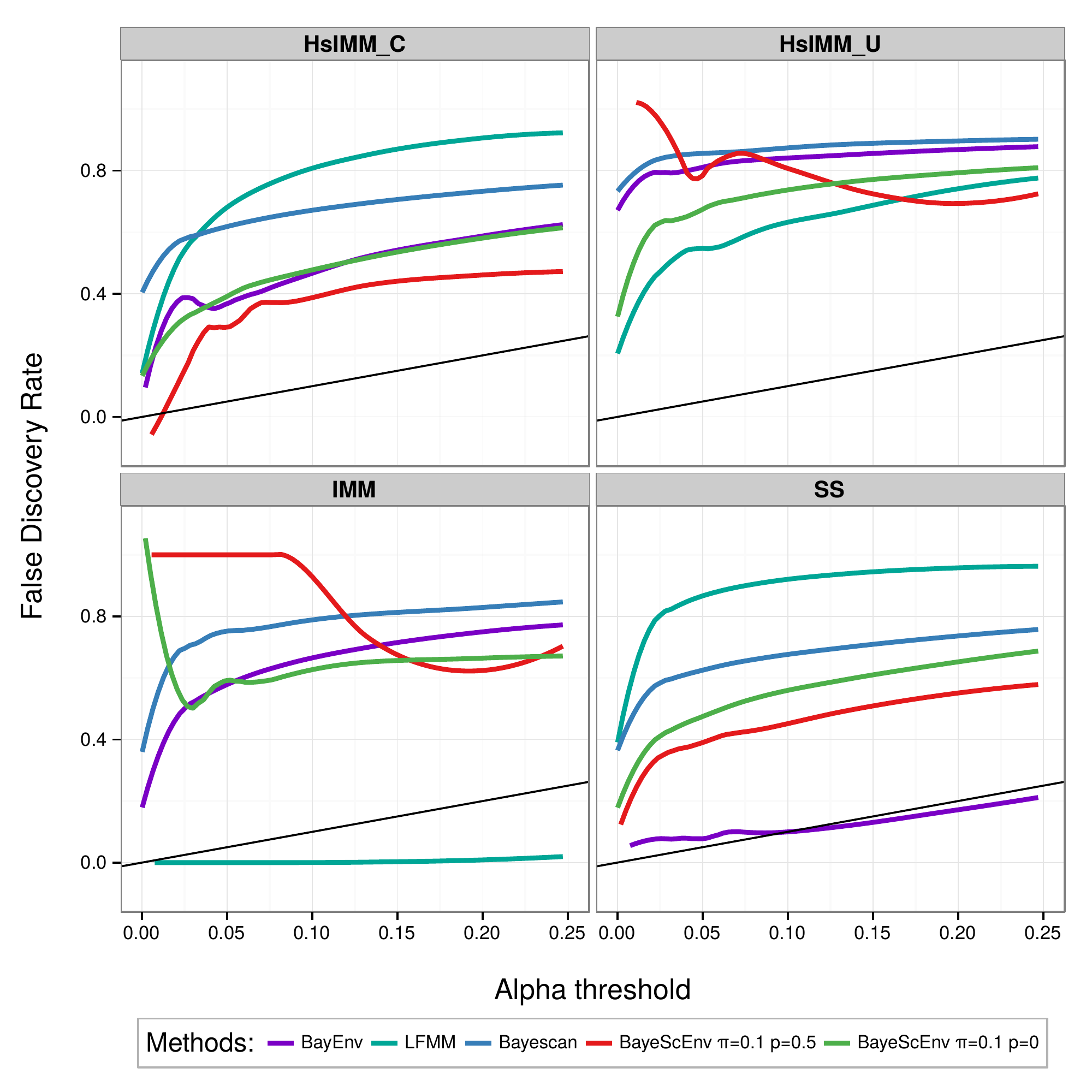}
   \caption{False Discovery Rate (FDR) against significance threshold $\alpha$ for \citet{de_villemereuil_genome_2014} polygenic 
 scenarios.}
   \label{supp_fig_ME_fdr}
 \end{center}
\end{minipage}
\hspace{\stretch{1}}
\begin{minipage}{0.4\textwidth}
 \begin{small}
 \paragraph{False Discovery Rate}
 When comparing the FDR to other methods, BayeScEnv performs relatively well. Especially, for $p=0.5$ (red line), its FDR can be the
 lowest (HsIMM-C) or second best (SS), but it can reach very high values under some scenarios (HsIMM-U and IMM). Surprisingly, the 
 parametrisation $p=0$ (green) is more stable across scenarios, whereas Bayenv (purple) and LFMM (turquoise) constantly ``switch'' 
 between best-or-so and poorest-or-so. Overall, BayeScEnv's FDRs with $p=0$ are lower than those of BayeScan's, at least for the 
 canonical threshold $\alpha=0.05$.
 
 Large values of FDR for small $\alpha$'s are due to a lack of power (see below), not to a high False Positive Rate.
 \end{small}
\end{minipage}
\hspace{\stretch{1}}
\end{figure}

\begin{figure}[H]
 \begin{minipage}{0.5\textwidth}
 \begin{center}
 \includegraphics[width=\textwidth]{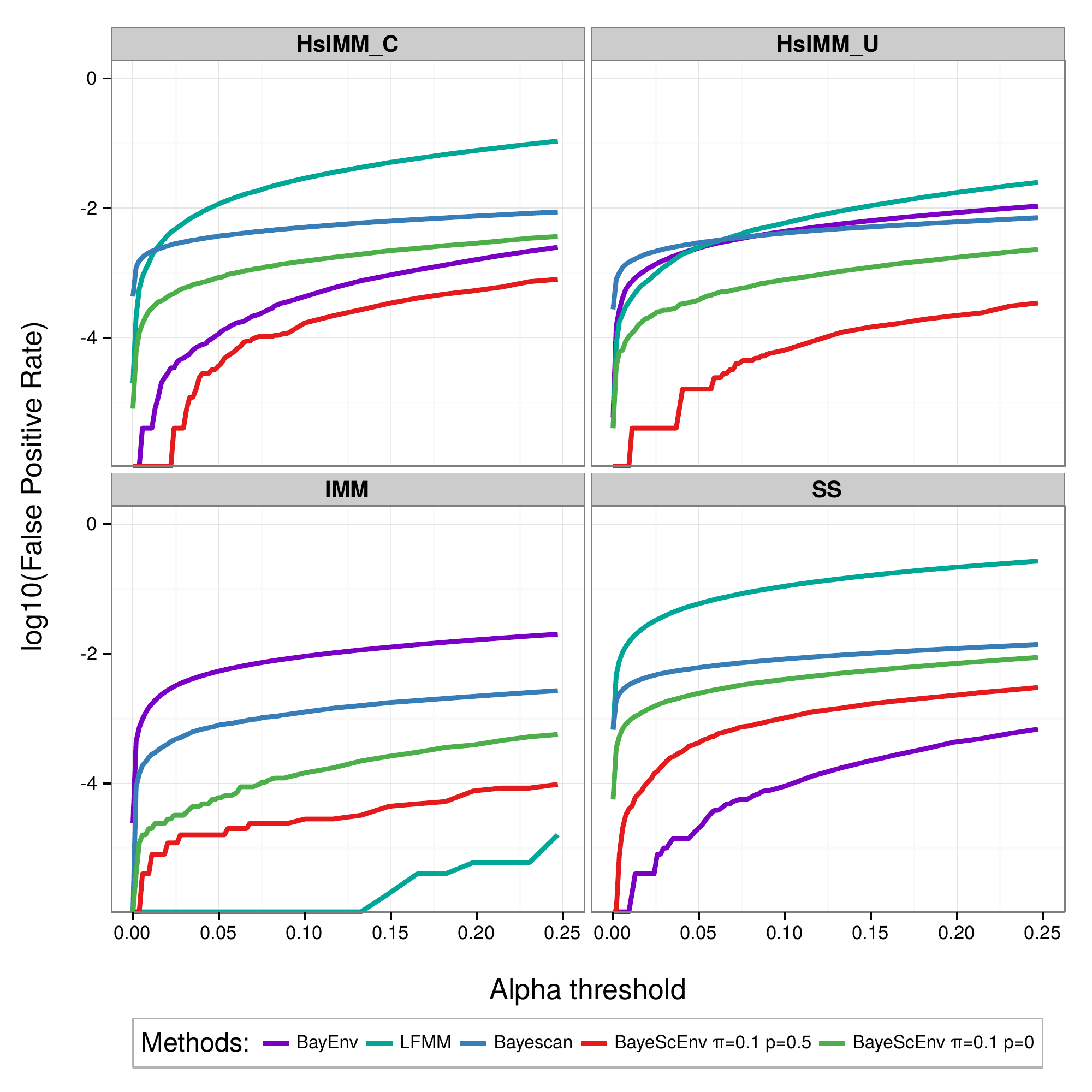}
 \caption{False Positive Rate (FPR) against significance threshold $\alpha$ for \citet{de_villemereuil_genome_2014} polygenic 
 scenarios.}
 \label{supp_fig_ME_fpr}
 \end{center}
\end{minipage}
\hspace{\stretch{1}}
\begin{minipage}{0.4\textwidth}
 \begin{small}
 \paragraph{False Positive Rate}
 FPRs are more predictable than FDRs regarding the $F$ model family: BayeScan (blue) is the most error-prone method, followed by 
 BayeScEnv with $p=0$ (green) while BayeScEnv with $p=0.5$ (red) is one of the most conservative methods. Interestingly, the
 FPRs of the $F$ model family are more stable than the FPRs of Bayenv (purple) and LFMM (turquoise), which vary greatly across 
 scenarios.
 \end{small}
\end{minipage}
\hspace{\stretch{1}}
\end{figure}

\begin{figure}[H]
 \begin{minipage}{0.5\textwidth}
 \begin{center}
 \includegraphics[width=\textwidth]{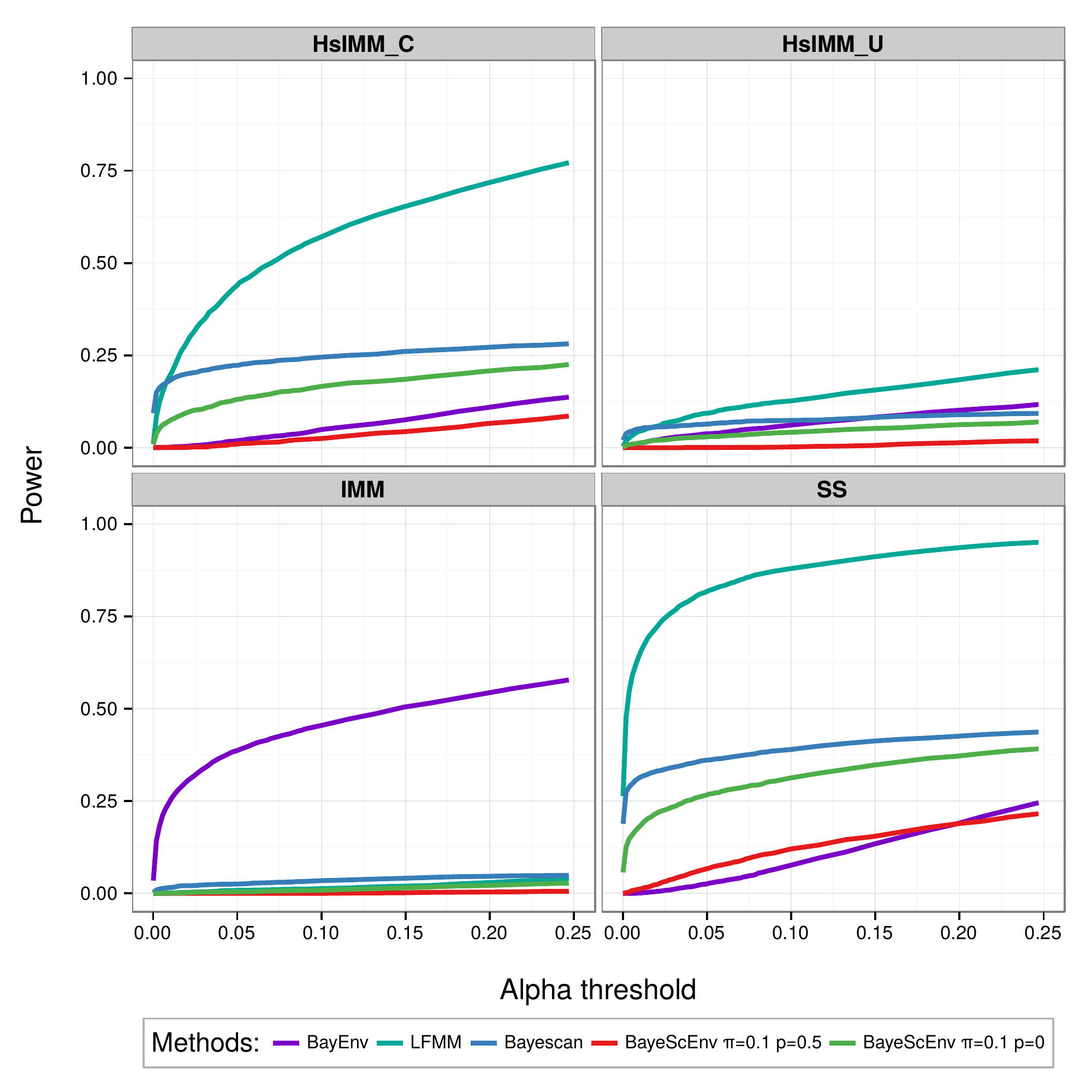}
 \caption{Power against significance threshold $\alpha$ for \citet{de_villemereuil_genome_2014} polygenic 
 scenarios.}
 \label{supp_fig_ME_power}
 \end{center}
\end{minipage}
\hspace{\stretch{1}}
\begin{minipage}{0.4\textwidth}
 \begin{small}
 \paragraph{Power}
 Overall power varies greatly across scenarios, HsIMM-U and IMM being the most difficult ones. As expected, the power of
 BayeScEnv (red and green) is always lower than the power of BayeScan (blue). However, the power of BayeScEnv with $p=0$ (green), 
 is always comparable to that of BayeScan's. BayeScEnv with $p=0.5$ (red) is always among the less powerful method. 
 For all scenarios, at least on of the environmental association methods (Bayenv (purple) and LFMM (turquoise)), has greater
 power than BayeScan and BayeScEnv.
 \end{small}
\end{minipage}
\hspace{\stretch{1}}
\end{figure}

\paragraph{Results}
The results are presented in Figs. \ref{supp_fig_ME_fdr}--\ref{supp_fig_ME_power} (below). They show that, even under very
difficult conditions, BayeScEnv inferences are fairly reliable. As expected, both the FPR and the power of BayeScEnv are lower than that of
BayeScan, resulting in a more conservative method overall. However, in scenarios with low power for all methods, BayeScEnv's lack
of power can drastically inflate its FDR (e.g. Fig. \ref{supp_fig_ME_fdr}, red line, IMM model). Interestingly, BayeScEnv with $p=0$
is more robust in this regard, since its power is generally much higher.\\
When compared to the other association methods (Bayenv and LFMM), BayeScEnv performed very well in ``clinal'' scenarios (HsIMM-C
and SS), but more poorly in the other scenarios. However when considering the canonical $\alpha=0.05$ threshold, BayeScEnv's FDR
was always lower than at least one of the association methods, except in the IMM scenario.

\newgeometry{left=0.7cm,right=0.7cm,bottom=3cm,top=2.5cm}

\restoregeometry

\subsection{Typical frequency patterns detected by BayeScEnv}

Consider the standardised environmental variable $X_{j}$ from which we derived the environmental differentiation using $E_{j}=|X_{j}|$.
As explained in the Discussion (see main text), we can distinguish three main patterns of population allele frequencies $f_{j}$ as a function of this environmental value $X_{j}$.

\begin{minipage}{0.3\textwidth}
 \begin{center}
 \includegraphics[width=\textwidth]{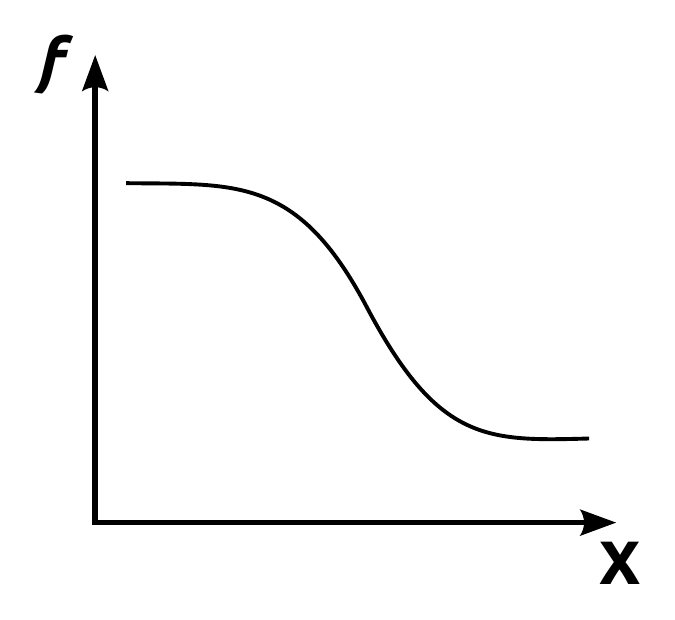}
 \end{center}
\end{minipage}
\begin{minipage}{0.69\textwidth}
 \begin{center}
    \begin{large} \textbf{Clinal scenario} \end{large}\\
    This is the canonical scenario, which was simulated in this study and most thoroughly investigated. It is also the
    scenario tested in all evaluations of environmental association methods.
 \end{center}
\end{minipage}

\begin{minipage}{0.3\textwidth}
 \begin{center}
  \includegraphics[width=\textwidth]{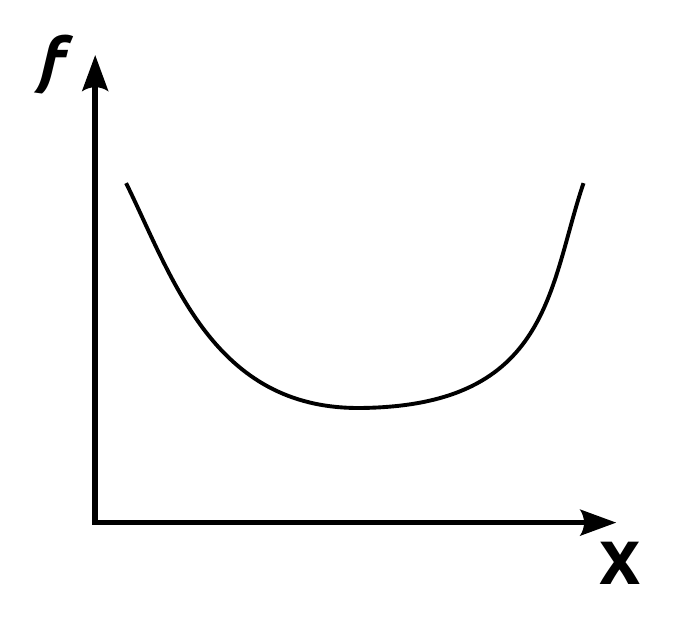}
 \end{center}
\end{minipage}
\begin{minipage}{0.69\textwidth}
 \begin{center}
    \begin{large} \textbf{``Plasticity'' scenario} \end{large}\\
    In this scenario, two populations with very different environmental values have similar allele frequencies. 
    It could arise in situations where the environmental variable used is a proxy for a true environmental variable that 
    has a non-monotonic relationship with the environmental variable used (e.g. very high or very low temperatures can both lead 
    to aridity). More interestingly, it is expected, for example, from genes responsible for phenotypic plasticity \citep{morris_gene_2014}.
 \end{center}
\end{minipage}

\begin{minipage}{0.3\textwidth}
 \begin{center}
  \includegraphics[width=\textwidth]{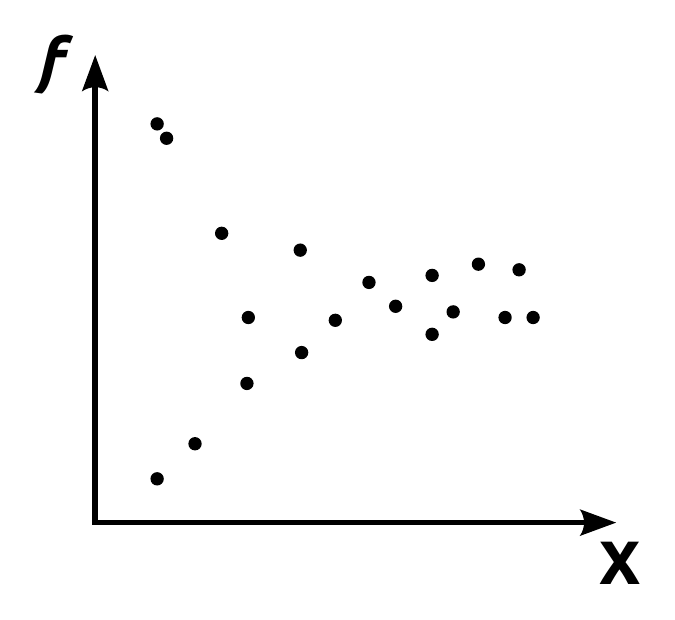}
 \end{center}
\end{minipage}
\begin{minipage}{0.69\textwidth}
 \begin{center}
    \begin{large} \textbf{Extreme frequencies scenario} \end{large}\\
    In this scenario, populations with similar environmental values have extremely different allele frequencies. 
    This scenario would lead to results that should in principle be interpreted as false
    positives. Note, however, that such a scenario could be explained by positive frequency-dependent selection
    triggered by the environmental variable.
 \end{center}
\end{minipage}

\end{document}